\documentclass[fleqn,usenatbib]{mnras}

\usepackage{newtxtext,newtxmath}
\usepackage[T1]{fontenc}

\DeclareRobustCommand{\VAN}[3]{#2}
\let\VANthebibliography\thebibliography
\def\thebibliography{\DeclareRobustCommand{\VAN}[3]{##3}\VANthebibliography}

\usepackage{graphicx}	% Including figure files
\usepackage{amsmath}	% Advanced maths commands
\usepackage{enumitem}
\usepackage{url}

\newcommand{\kapSeven}{$1.9\times 10^{-2}$}
\newcommand{\kapSix}{$1.3\times 10^{-2}$}
\newcommand{\kapThree}{$4.9\times 10^{-3}$}

\title[Dust Opacities and Temperature of HH 212 Disk]{Inferring (Sub)millimeter Dust Opacities and Temperature Structure in Edge-on Protostellar Disks From Resolved Multi-Wavelength Continuum Observations: The Case of the HH 212 Disk}

\author[Z.-Y. D. Lin et al.]{
Zhe-Yu Daniel Lin,$^{1,2,3}$\thanks{E-mail: zdl3gk@virginia.edu}
Chin-Fei Lee,$^{2,4}$
Zhi-Yun Li,$^{1}$
John J. Tobin,$^{5}$
and Neal J. Turner$^{6}$
\\
$^{1}$Department of Astronomy, University of Virginia, Charlottesville, Virginia 22904, USA\\
$^{2}$Academia Sinica Institute of Astronomy and Astrophysics, P.O. Box 23-141, Taipei 106, Taiwan\\
$^{3}$Department of Physics, National Chung Cheng University, No.168, Sec. 1, University Rd., Minhsiung, Chiayi 621301, Taiwan \\
$^{4}$Graduate Institute of Astronomy and Astrophysics, National Taiwan University, No. 1, Sec. 4, Roosevelt Rd., Taipei 10617, Taiwan\\
$^{5}$National Radio Astronomy Observatory, 520 Edgemont Road, Charlottesville, Virginia 22903, USA \\
$^{6}$Jet Propulsion Laboratory, California Institute of Technology, Pasadena, California 91109, USA
}

\date{Accepted XXX. Received YYY; in original form ZZZ}

\pubyear{2020}

\begin{document}
\label{firstpage}
\pagerange{\pageref{firstpage}--\pageref{lastpage}}
\maketitle

\begin{abstract}

(Sub)millimeter dust opacities are required for converting the observable dust continuum emission to the mass, but their values have long been uncertain, especially in disks around young stellar objects. We propose a method to constrain the opacity $\kappa_\nu$ in edge-on disks from a characteristic optical depth $\tau_{0,\nu}$, the density $\rho_0$ and radius $R_0$ at the disk outer edge through $\kappa_\nu=\tau_{0,\nu}/(\rho_0 R_0)$ where $\tau_{0,\nu}$ is inferred from the shape of the observed flux along the major axis, $\rho_0$ from gravitational stability considerations, and $R_0$ from direct imaging. We applied the 1D semi-analytical model to the embedded, Class 0, HH 212 disk, which has high-resolution data in ALMA Band 9, 7, 6, and 3 and VLA Ka band ($\lambda$=0.43, 0.85, 1.3, 2.9, and 9.1 mm). The modeling is extended to 2D through \textsc{RADMC-3D} radiative transfer calculations. We find a dust opacity of $\kappa_\nu \approx $ \kapSeven, \kapSix, and \kapThree cm$^2$ per gram of gas and dust for ALMA Bands 7, 6, and 3, respectively with uncertainties dependent on the adopted stellar mass. The inferred opacities lend support to the widely used prescription $\kappa_\lambda=2.3\times 10^{-2} (1.3 {\rm mm}/\lambda)$ cm$^2$~g$^{-1}$ 
%advocated by \cite{Beckwith1990}
. We inferred a temperature of $\sim 45$~K at the disk outer edge which increases radially inward. It is well above the sublimation temperatures of ices such as CO and N$_2$, which supports the notion that the disk chemistry cannot be completely inherited from the protostellar envelope.
\end{abstract}


\section{Introduction}
\label{sec:introduction}

% what greater problem does this project address? 
% - how do disks form? 
% - provide understanding for connecting envelope to disk and disk evolution itself
% many high angular resolution images, survey of Class 0/I sources 
% - 
% describe past HH 212 data
% - jet, envelope, resolved disk, COMs, polarization from disk, line polarization in jet 
% - resolved disk in the vertical direction at Band 7
% peer into the disk

Disks around young stellar objects are the birthplaces of planets. The formation and evolution of the planets depend critically on the mass of the dust and gas in the disk. Although the disk mass can be obtained from observations of molecules in some cases, especially through HD \citep[e.g.][]{Bergin2018}, it is mostly determined from (sub)millimeter dust continuum observations. To convert the observed continuum flux to the mass, a dust opacity is required. The most widely used opacity (relative to gas mass) in the literature is the prescription \citep{Beckwith1990}:
\begin{equation}
    \kappa_\lambda = 2.3\times 10^{-2} \left(\frac{1.3\ {\rm mm}}{\lambda}\right)\ {\rm cm}^2\ \rm{g}^{-1} \text{ .}
    \label{beckwith_op}
\end{equation} 
If converted to dust opacity with a standard gas-to-dust mass ratio of 100, it is well known that this opacity is well above the values predicted for diffuse interstellar medium \citep{Draine1984}. It is generally accepted that the enhanced opacity comes from grain growth in the disk (\citealt{Ossenkopf1994}; see \citealt{Testi2014} for a review), but the detailed properties of the grains (including their shape, composition and size) are uncertain, making quantitative predictions from first principles difficult. Empirical constraints of the dust opacities in disks are therefore highly desirable. It is one of the main goals of our investigation. 

%{why HH 212} basic idea, then why HH 212

One way to constrain the dust opacity $\kappa_\nu$ at a given frequency in a disk is through the optical depth $\tau_\nu=\kappa_\nu\Sigma$, where $\Sigma$ is the column density along a line of sight through the disk. To determine the optical depth, it is desirable to have a disk that shows a clear transition from optically thin sight lines to optically thick ones. The column density can be constrained from dynamical considerations, especially gravitational stability. This is where the HH 212 mms protostellar disk comes in.

HH 212 mms is a well studied, nearly edge-on, protostellar system with a spectacular, symmetric bipolar jet powered by the central Class 0 protostar IRAS 05413-0104 \citep{Chini1997, Zinnecker1992, Zinnecker1998}. It is located in the L1630 cloud in Orion at a distance of about 400 pc \citep{Kounkel2017}. Images of continuum, CO, and HCO+ reveal a flattened, massive envelope ($\sim 0.06 - 0.1 M_{\odot}$) around the central source, leaving little doubt that it is a very young system that is still in active formation \citep{Lee2006, Lee2008, Lee2014}. One of the most important features of the HH 212 system that makes it ideal for our purposes of constraining dust opacities is its nearly edge-on orientation and the dark lane along the disk equatorial plane observed by \cite{Lee2017_lane} in the Atacama Large Millimeter/submillimeter Array (ALMA) Band 7 ($\lambda=0.85$~mm, 0.02\arcsec resolution), which demonstrated unequivocally that the disk is optically thick along sight lines close to the center and becomes optically thin near the outer edge. As we will show in detail in the paper, this transition provides a quantitative handle on the optical depth. 

Another useful feature of the HH 212 system is that it has a rather low stellar mass, estimated to be $\sim 0.2 - 0.25 M_{\odot}$ \citep{Lee2014, Lee2017_com}. The disk, on the other hand, is rather bright in dust continuum \citep{Lee2017_lane}, which is indicative of a relatively large disk mass. The combination of a relatively small stellar mass and a relatively large disk mass makes the system prone to gravitational instability. Indeed, \cite{Tobin2020} estimated a Toomre parameter $Q$ that lies between 1 and 2.5, indicating that the disk is marginally gravitational unstable, with potential development of spiral arms that can help redistribute angular momentum in the disk and keep the $Q$ parameter somewhat above unity \citep[e.g.][]{Bertin1999, Kratter2010, Kratter2016, Takahashi2016}. Although such spirals would be difficult to detect in HH 212 mms because of its edge-on orientation, they are observed in a comparable protostellar disk, HH 111 mms \citep{Lee2020}, which the authors argued is driven gravitationally unstable by fast accretion from the protostellar envelope, as shown in the numerical simulations of, e.g., \cite{Tomida2017}. It is reasonable to expect the HH 212 disk to be marginally gravitationally unstable as well, since it is also fed by a massive infalling envelope and its stellar mass ($\sim 0.25$ M$_\odot$) is much lower than that of HH 111 mms ($\sim 1.5$~M$_\odot$; \citealt{Lee2020}), even though its disk is less bright at $0.85$~mm, but only by a factor of $\sim 3.5$ \citep{Lee2018_pol}. The degree of gravitational stability, as characterized by the Toomre $Q$ parameter, provides a handle on the disk column density, which is required for the dust opacity determination from the optical depth. 

Just as importantly, the HH 212 disk is currently one of the best resolved Class 0 disk at multiple wavelengths. Besides existing data in ALMA Band 7 ($\lambda=0.85$~mm) and the Very Large Array (VLA) Band Ka (9.1~mm; \citealt{Tobin2020}), we will present new high resolution data in ALMA Band 9 (0.43~mm), Band 6 (1.3~mm) and Band 3 (2.9~mm). Multi-wavelength data are important not only for constraining the dust opacities at these respective wavelengths but also for inferring the temperature distribution of this edge-on disk along its midplane because the disk is expected to have different opacities at different wavelengths, with their dust emission coming from different depths into the disk along the line of sight and thus probing the temperatures over a range of radii. The temperature determination is important because it directly affects not only the disk structure (through, e.g., the scale height) but also its chemistry, particularly ices of low sublimation temperatures such as CO and N$_2$ \citep[e.g.][]{Pontoppidan2014, vantHoff2018, vantHoff2020}. 

The rest of the paper is organized as follows. We will first present new high resolution dust continuum data of the HH 212 disk in ALMA Band 9, 6 and 3, together with the existing data in ALMA Band 7 and VLA Band Ka in Section \ref{sec:obs}, with a focus on the distributions of their brightness temperatures along the major axis. In Section \ref{sec:semianalytical}, we will illustrate how to use the brightness temperature profile to constrain the dust opacity and determine the temperature distribution along the disk midplane through a simple, one-dimensional (1D) semi-analytical model and apply the model to the multi-wavelength data set of the HH 212 disk. In Section \ref{sec:axisymmetric}, we extend the HH 212 mms modeling through \textsc{RADMC-3D} radiative transfer calculations. We find that the (sub)-millimeter dust opacities in the HH 212 disk are consistent with the widely used values advocated by \citep{Beckwith1990} and that the HH 212 disk is warmer than 
$\sim 45$~K everywhere, including the midplane. The implications of these and other results are discussed in Section \ref{sec:discussion}. Our main conclusions are summarized in Section \ref{sec:conclusions}. 

%\cite{Lee2017_com} detected complex organic molecules in the disk atmosphere and together with the high angular resolution images of HCO+, they found that the inner envelop and disk can be described by a simple model of a rotating and infalling system with conserved angular momentum. As a result, the system can be separated into two distinct regions. The region within 

% ================================================
\section{Multi-Wavelength Dust Continuum Observations of HH 212 mms} \label{sec:obs}

HH 212 mms (05:43:51.4107, -01:02:53.167, ICRS) was observed by ALMA in Bands 3, 6, 7, and 9 and by the VLA at Band Ka. Table \ref{table:ObservationLog} lists the observation logs and Table \ref{table:calibrator} records the calibrators. The raw visibility data of each ALMA band were calibrated using the standard reductions scripts, which uses tasks in Common Astronomy Software Applications (CASA; \cite{CASA_McMullin}). Continuum images were created from the calibrated measurement sets with the CLEAN task and channels with line emission were ignored. The observations are summarized in Tables~\ref{table:ObservationLog} and \ref{table:calibrator}. 

ALMA Band 9 ($0.43$~mm) was observed on 2015 July 26 with 2 execution blocks. The total on-source time is 44 minutes using 35 to 41 antennas with baselines of 15 to 1574.4 meters. The correlator observes a total bandwidth of 8 GHz. The continuum image centers on \(\sim\)691.5 GHz and uses a  robust parameter of +0.5. The resulting beam has a FWHM of \(0.081\arcsec\) $\times$ \(0.061\arcsec\) and PA of 66.5$^{\circ}$.    

The ALMA Band 7 continuum ($0.85$ mm) was first published in \cite{Lee2017_lane}. The image combined four different executions, two of them on 2015 August 29 and one execution on 2015 November 05 and on 2015 December 03. The total baseline ranges from 15 to 16196m, while the total observation time is 148 minutes. The resulting beam has a FWHM of \(0.020\arcsec\) $\times$ \(0.018\arcsec\) and PA of $-65.6^{\circ}$.

ALMA Band 6 is observed on 2017 October 04. The observation time is 19 minutes using 46 antennas to span baselines of 41 to 14969m. The correlator covers a bandwidth of 7.5 GHz with channel widths of 976.5 KHz. The continuum image centers on \(\sim\)225.9 GHz (1.3 mm) and uses robust parameter of +0.5. The resulting beam has a FWHM of \(0.026\arcsec\) $\times$ \(0.021\arcsec\) and PA of 60.3$^{\circ}$.

The ALMA Band 3 observation was executed on 2017 October 05. The on-source time is 50 minutes and 46 of ALMA 12m antennas were used with baselines from 41 to 14969m. The correlator covers a bandwidth of 7.5 GHz in frequency with channel widths of 976.5 KHz. The continuum image centers on \(\sim\)105.0 GHz with Briggs weighting of a robust parameter of -0.5 when cleaning. The resulting beam has a FWHM of \(0.044\arcsec\) $\times$ \(0.029\arcsec\) and a PA of -84.5$^{\circ}$.

Band Ka was observed by the VLA as part of the VLA/ALMA Nascent Disk and Multiplicity (VANDAM) survey for Orion protostars. HH 212 mms was observed on 2016 October 22 with the A configuration with a time-on-source of $\sim$1 hr. The correlator covers a bandwidth of 8 GHz, with 4 GHz centered at $\sim$37 GHz and another 4 GHz centered at 29 GHz. The continuum was imaged using natural weighting and multifrequency synthesis with \textit{nterms}=2 across both basebands. The resulting beam has a FWHM of $0.062 \arcsec$ $\times$ $0.055 \arcsec$ and a PA of $-80.8^{\circ}$.

% table for observations 
\begin{table*}
	\centering
	\caption{Observation logs for each band}
	\begin{tabular}{ccccccccc} %alignment for each
		\hline
		Band & Wavelength & Time & Baselines & Beam & Beam PA & Noise & Peak $I_{\nu}$ & Total Flux \\
		& (mm) & (min) & (m) & (mas) & ($^{\circ}$) & (mJy beam$^{-1}$) & (mJy beam$^{-1}$) & (mJy) \\
		\hline
		9 & 0.434 & 44.18 & 15$\sim$1574.4  & 81x61     & 66.5  & 3.0   & 100.9 & 1.157 $\times 10^{3}$ \\
		7 & 0.851 & 144 & 15$\sim$16196     & 20x18     & -65.6 & 0.08  & 3.34 & 151.8 \\
		6 & 1.33 & 18.73 & 41$\sim$14949    & 26x21     & 60.3  & 0.022 & 2.32 & 59.26 \\
		3 & 2.85 & 49.62 & 41$\sim$14969    & 44x29     & -84.5 & 0.018 & 1.30 & 10.05 \\
		Ka & 9.10 & 57.67 &  793$\sim$34425     & 62x55     & -80.8 & 9.1 $\times 10^{-3}$  & 0.258 & 0.508 \\
		\hline
	\end{tabular}
		\label{table:ObservationLog}
\end{table*}
% Total Flux [Jy/beam]: 1.15697556e+00 1.51842637e-01 5.92644113e-02 1.00528245e-02 5.07975083e-04

% table for calibrators
\begin{table}
%	\label{table:calibrator}
	\centering
	\caption{Calibrators for each band. }
	\begin{tabular}{ccccc} %alignment for each
		\hline
		Band & Date & Bandpass& Flux & Phase\\
		& & Calibrator & Calibrator & Calibrator \\
		\hline
		9 &  2015-07-26 & J0522-3627 & J0423-013 & J0607-0834\\
		7 & 2015-08-29 & J0607-0834 & J0423-013 & J0552+0313 \\
		 & 2015-11-05 & J0423-0120 & J0423-0120 & J0541-0211 \\
		 & 2015-12-03 & J0510+1800 & J0423-0120 & J0541-0211 \\
		6 & 2017-10-04 & J0423-0120 & J0423-0120 & J0541-0211\\
		3 & 2017-10-05 & J0423-0120 & J0423-0120 & J0541-0211\\
		Ka & 2016-10-22 & J0319+4130 & J0137+3309 & J0552+0313 \\
		\hline
	\end{tabular}
		\label{table:calibrator}
\end{table}

% ================================================

\subsection{Images}

\begin{figure}
    \centering
    \includegraphics[width=\columnwidth]{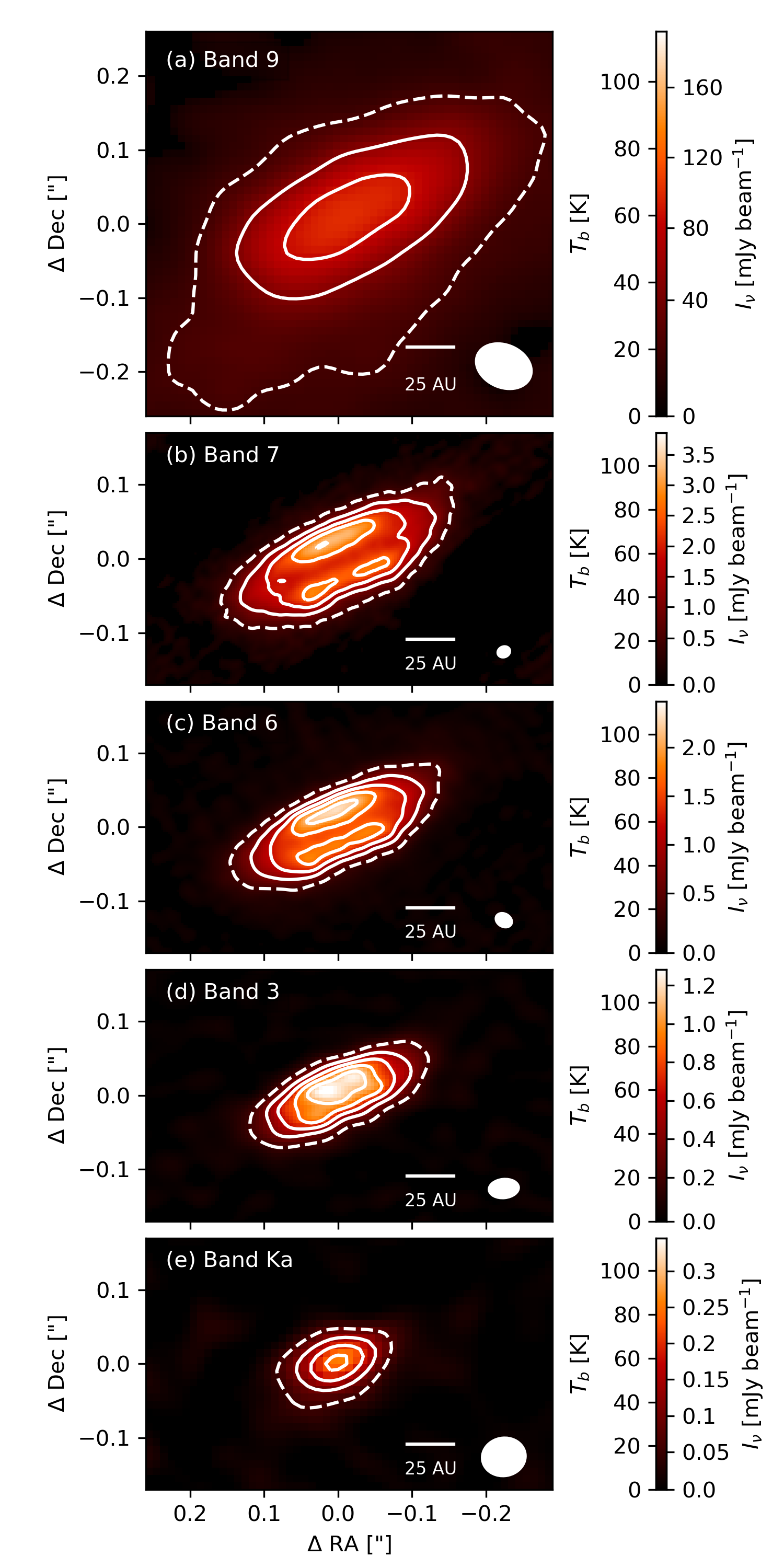}
    \caption{
        The color maps depict the brightness temperature distributions of the dust continuum emission at ALMA Bands 9 to 3 and VLA Ka band. The color bar shows both the brightness temperature in Kelvin and the flux density in mJy beam$^{-1}$. The range of the brightness temperature is from 0 to 115 K and is the same for all 5 images. The contours are brightness temperature levels starting from 20K (the dashed contour) in steps of 20. The white filled ellipses in the lower right of each panel is the FWHM of the corresponding beam. 
           }
    \label{fig:realImages}
\end{figure}

Fig. \ref{fig:realImages} shows the continuum images at the five different wavelengths plotted in flux density $I_{\nu}$ in mJy beam$^{-1}$ and brightness temperature $T_{b}$ in Kelvin. We use the full Planck function to convert the observed flux into the brightness temperature at each waveband, since the Rayleigh-Jeans approximation deviates significantly at the shorter wavelengths. 
All images show elongated structures which represent the circumstellar disk described in \cite{Lee2017_lane} with the major axis of the disk perpendicular to the jet (oriented $23^{\circ}$ East-of-North in the plane of sky; \citealt{Lee2017_jet}). The jet is red-shifted to the South. The emission seen in Band 9 is the most extended compared to the longer wavelengths. However, no detailed features are seen, since the beam size is the largest and structures are likely not as well resolved. The Band 7 image has a dark lane along the major axis that is sandwiched by two brighter layers and this was interpreted as coming from a vertical temperature gradient. The asymmetry of the bright layers is due to inclination where the near-side of the disk (marked by the red-shift jet) is in the southwest direction. The dark lane seen in Band 7 is also clearly detected in Band 6. At Band 3, the dark lane is no longer as prominent and only an asymmetry along the disk minor axis is observed. Detailed features no longer exist in Band Ka and the disk is only a couple beams wide. The extent of the emission is largest at the shortest wavelength (Band 9; partly because of its lower resolution) and decreases as the wavelength increases. Qualitatively, this trend makes physical sense, since dust opacity decreases with increasing wavelength, which leads to a decrease in the total optical depth ($\tau$) along the line of sight and thus the area in the sky where the total optical depth is above unity. 

\begin{figure*}
    \centering
    \includegraphics[width=\textwidth]{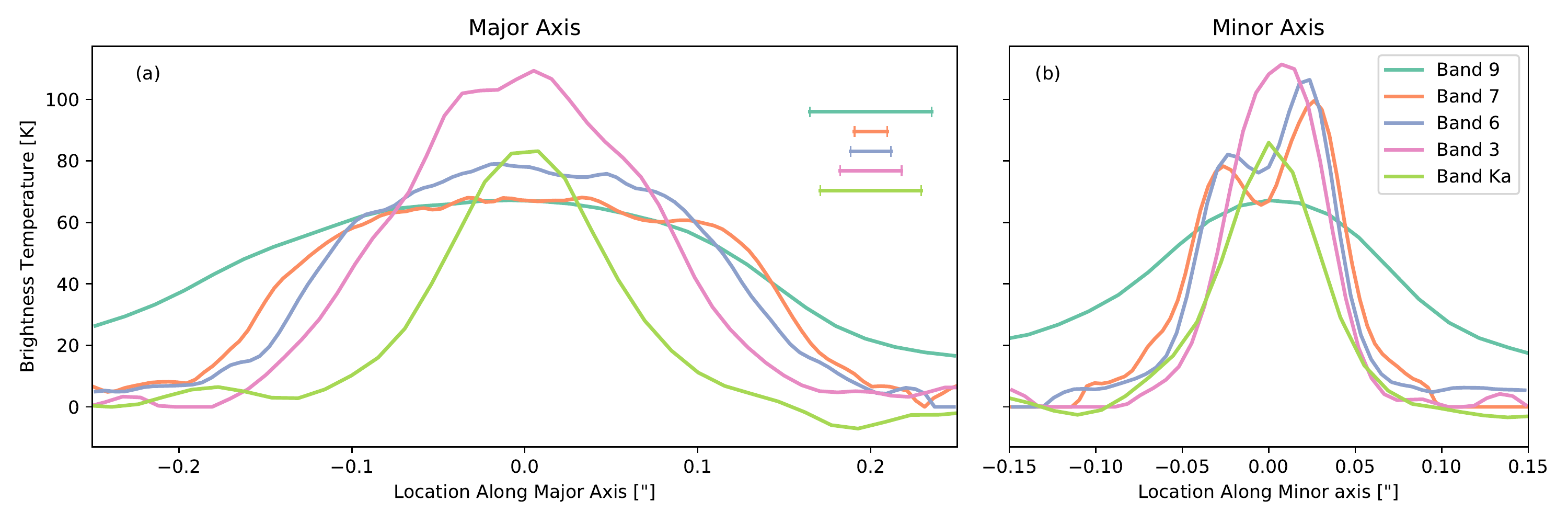}
    \caption{
        The cuts a long the major (Panel a) and minor (Panel b) axes in brightness temperature. The length of the geometric mean of the FWHM of the beam is shown in the upper right corner of Panel a. 
            }
    \label{fig:tbcut}
\end{figure*}

For a more quantitative comparison of the brightness distributions at different wavelengths, we plot the brightness temperature along the major and minor axis of the disk in Fig. \ref{fig:tbcut}. We take the peak of the continuum at the VLA Ka band to be the location of the central source (see Appendix \ref{sec:diskcenter} for details). There are three notable features of the profile along the major axis with increasing wavelength (Fig. \ref{fig:tbcut}a). First, the width of the profile (e.g., the full width at half maximum) decreases as expected based on the images seen above. Second, the peak of the profile increases up until Band 3 and decreases at Band Ka. Third, the Band Ka profile is more Gaussian-like or "peaky," while in comparison, Bands 7, 6, and 3 appear to be ``boxy." Explicitly, the central regions roughly plateau at a constant brightness temperature and drops dramatically outside the plateau with Bands 7 and 6 as the most obvious cases. Band 9 roughly shares the same plateau as Band 7 though with more extended "wings."

Along the disk minor axis (Fig. \ref{fig:tbcut}b), the brightness temperature cuts also have a decreasing profile width with increasing wavelength. Furthermore, the peak of the profile increases with increasing wavelength except at Band Ka which is reminiscent of the major axis cuts. The central dip in Band 6 and 7 profiles are where the dark lane is located with a contrast of the brightest peak to the dip of around $\sim 30\%$ for both bands. At Band 3, the dip no longer exists and only has a single peak shifted to the far-side of the disk. The Band Ka profile resembles a centrally peaked Gaussian profile. 

\subsection{Spectral Index} \label{ssec:spectral_index}

\begin{figure}
    \centering
    \includegraphics[width=\columnwidth]{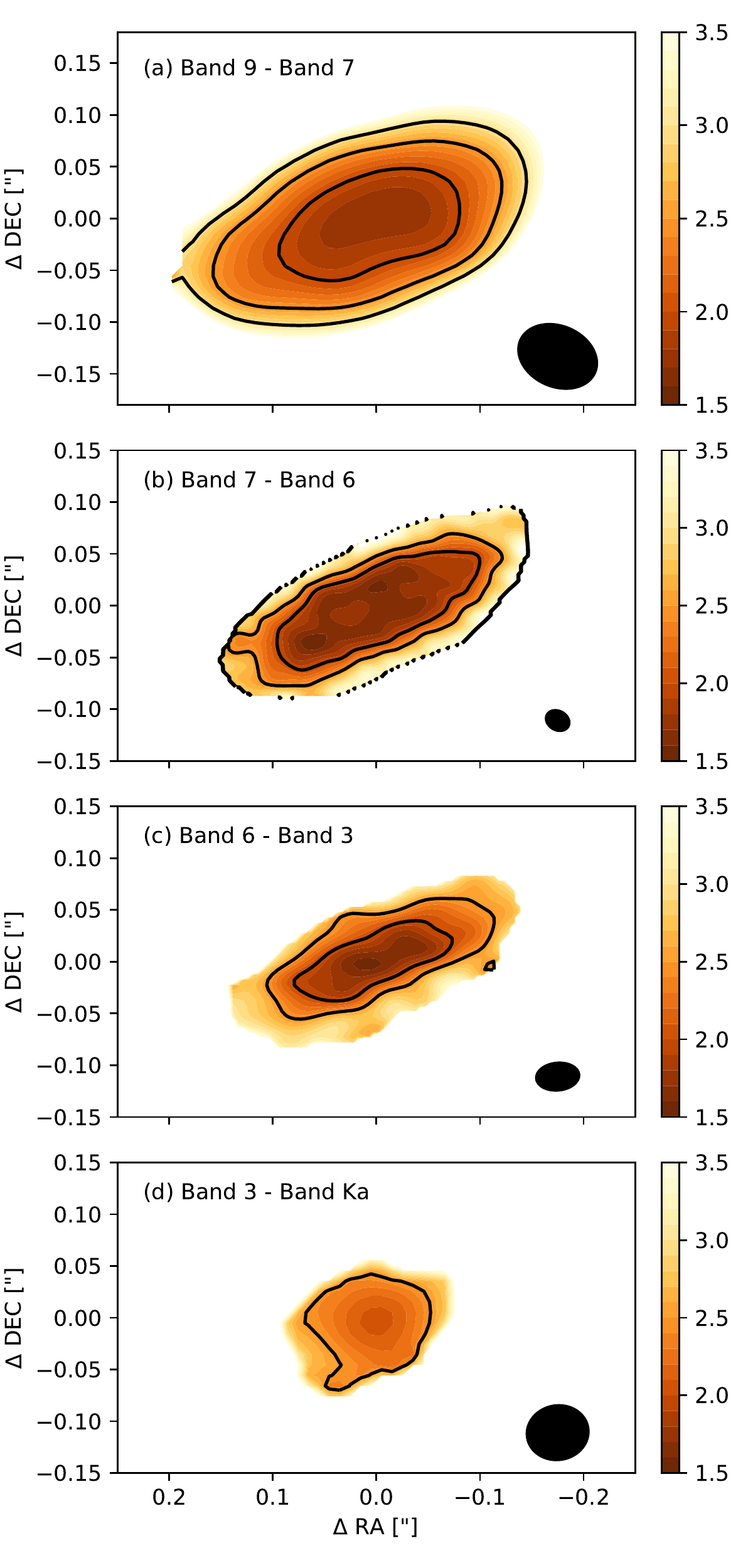}
    \caption{
        The spectral index map between consecutive wavelengths after adopting a common resolution. The black ellipses in each panel are the corresponding beam. The black contours show where $\alpha=2$, 2.5, and 3. 
            }
    \label{fig:multialpha}
\end{figure}

We show the spectral index images between consecutive wavelengths in Fig. \ref{fig:multialpha}, defined by $\alpha \equiv \partial \ln I_{\nu} / \partial \ln \nu$ where $I_{\nu}$ is the intensity and $\nu$ is the frequency. Between each pair, the image with the higher resolution was convolved by the CASA command \texttt{imsmooth} to the same resolution as the image of lower resolution. We ignore the regions with flux densities less than 5 times the noise.

All of the spectral index images have mostly $\alpha < 3$. For the first three pairs (from Bands 9, 7, 6, and 3; Fig. \ref{fig:multialpha}a, b, c), the inner regions of the disk has $\alpha < 2$. Only the Band 3 and Band Ka pair does not have $\alpha <2$ (Fig. \ref{fig:multialpha}d). 

The relative low values of $\alpha$ is interesting because for an optically thin ($\tau \ll 1$) slab of isothermally emitting dust in the Rayleigh-Jeans limit, one expects $\alpha = 2 + \beta$ where $\beta$ is the opacity index defined by $\kappa_{\nu} \propto \nu^{\beta}$ \citep[e.g.][]{Draine2006}. If the object is not optically thin, $\alpha$ decreases and in the optically thick limit ($\tau \gg 1$), $\alpha = 2$ which is simply the frequency dependence of the Rayleigh-Jeans tail. In this case, $\beta$ cannot be directly obtained from $\alpha$. Several mechanisms can cause $\alpha < 2$, such as deviation from the Rayleigh-Jeans limit, dust scattering \citep{Zhu2019,Liu2019}, temperature gradient \citep{Galvan2018}, or a combination of these effects \citep{Carrasco2019, Sierra2020, Lin2020_specpol}. We will return to a discussion of the spectral index $\alpha$ in Section~\ref{ssec:2dresults} and the opacity index $\beta$ inferred based on modeling of the multi-wavelength dust continuum data in Section~\ref{ssec:opacity}.  

% ================================================
\section{Inferring Dust Opacity and Midplane Temperature: 1D Semi-Analytical Model} \label{sec:semianalytical}
% derivation 
% explain the physical insight: the peak, the plateau, the width, the cutoff
% fit band 7 and band 6

As discussed above and shown in Fig.~\ref{fig:tbcut}a, there are two well-defined trends among the three best spatially resolved wavebands (Band 7, 6 and 3), namely, the brightness temperature distribution along the major axis of the nearly edge-on disk becomes narrower and peaks at a higher value as the wavelength increases. In this section, we will develop a simple, approximate, semi-analytical model to understand these trends physically and to illustrate how to use the multi-wavelength continuum data to quantitatively constrain the radial temperature distribution along the disk midplane and the dust opacities at the observing wavelengths. 

\subsection{Qualitative Discussion}
\label{ssec:qualitative}

% qualitative ideas to guide the model and discussion, so that people know what to come and would not get lost early on. 
Qualitatively, it is straightforward to understand why the observed brightness temperature distribution $T_b$ along the major axis in the plane of the sky can yield information about the {\it radial} distribution of the disk midplane temperature. The key lies in the variable optical depth $\tau'$ between a location inside the disk and the observer (see Eq.~(\ref{eq:optdepth_p3}) below). For a nearly edge-on disk such as HH 212 mms, the line of sight to the central star is expected to be optically thick (i.e., the total optical depth $\tau \gg 1$). Most of the dust emission along this sight line comes from near the $\tau'=1$ surface (whose distance from the central star is denoted by radius $R_1$). The temperature at the $\tau'=1$ surface, $T_1$, should be close to the observed brightness temperature along the same sight line, according to the Eddington-Barbier relation in radiative transfer\footnote{The Eddington-Barbier relation holds when the opacity is dominated by absorption rather than scattering, an assumption that we will adopt for simplicity for the models to be presented in the main text. This simplification is motivated by the fact that, in the deeply embedded protostellar disk HH 212 mms which is still actively accreting from a massive envelope, there is no firm evidence for grain growth to mm/cm sizes; indeed, dust settling is not required for modeling the disk vertical structure (see Section \ref{ssec:NoSettling} below), indicating that the grains have yet to grow large enough to settle towards the midplane, unlike the more evolved disks such as HL Tau (e.g., \citealt{Pinte2016}). We will briefly explore the effects of scattering due to large grains in Appendix \ref{sec:scattering}.}. If the radius $R_1$ can be determined, the temperature as a function of radius, $T_1(R_1)$, would be ascertained. 

The radius $R_1$ is in fact related to the half-width $W$ of the observed profile of the brightness temperature along the major axis. The profile is determined to a large extent by the variation of the total optical depth $\tau$, which is much larger than unity towards the center and close to zero towards the edge. The half-width $W$ is approximately the distance from the center where $\tau \sim 1$: the brightness temperature should stay relatively constant along optically thick sight lines interior to $W$ and drop off quickly outside $W$ where the sight line becomes increasingly optically thin towards the edge. Clearly, at any given wavelength, the half-width $W$ depends on the dust opacity $\kappa_\nu$: the smaller the $\kappa_\nu$ is, the closer to the center the transition from optically thick to optically thin occurs, with a smaller 
$W$ (i.e., narrower $T_b$ profile). Similarly, the radius $R_1$ is also controlled by $\kappa_\nu$: the smaller the $\kappa_\nu$ becomes, the closer the $\tau'=1$ surface is from the star, with a smaller $R_1$. This common dependence on dust opacity yields a relation between $R_1$ and $W$, which we will quantify below through our semi-analytical model (see Eq.~\ref{eq:rbrelation} below). It enables us to determine $R_1$ from the observed $T_b$ profile for each waveband and thus the radial temperature distribution $T_1(R_1)$ from observations at different wavelengths that have different dust opacities. 

The dust opacity $\kappa_\nu$ can be constrained from the optical depth. For a qualitative discussion here, we can characterize the optical depth of an edge-on disk by $\tau_{0,\nu}\equiv \rho_0 R_0 \kappa_\nu$, where $R_0$ and $\rho_0$ are the radius and mass density at the disk outer edge. This characteristic optical depth controls to a large extent the shape of the brightness temperature profile along the major axis: the larger the value of $\tau_{0,\nu}$ is, the more boxy the profile becomes (as we demonstrate quantitatively below in Fig.~\ref{fig:semiana_num}). In order to produce a relatively boxy profile observed in, e.g., Band 7 (see Fig.~\ref{fig:tbcut}a, brown curve), the characteristic optical depth $\tau_{0,\nu}$ must be of order unity in order for the transition from optically thick to optically thin to occur along a sight line relatively close to the disk edge (its  value will be obtained through quantitative modeling). Since the disk radius $R_0$ can be estimated directly from the dust continuum images, we should be able to constrain the dust opacity 
\begin{equation}
    \kappa_\nu=\dfrac{\tau_{0,\nu}}{\rho_0 R_0}
    \label{eq:op1}
\end{equation}
if the mass density $\rho_0$ can be constrained. 

The mass density can be constrained from stability considerations. As discussed in the introduction (Section~\ref{sec:introduction}), the HH 212 disk is likely marginally gravitationally unstable, with Toomre $Q$ parameter of order unity \citep{Kratter2016, Tobin2020}. To relate the mass density $\rho_0$ to the Toomre $Q$ parameter, we note that  \citep{Toomre1964}:
\begin{equation} \label{eq:toomreQ}
    Q = \dfrac{c_{s} \Omega }{\pi G \Sigma}
\end{equation}
where $c_{s}$ is the isothermal sound speed, $\Omega$ is the Keplerian frequency, and $\Sigma$ is the surface density. We assume the gas and dust are well-mixed. The Keplerian frequency depends on the central star $M_{*}$ through $\Omega = \sqrt{GM_{*}/R^{3}}$ where $R$ is the cylindrical radius. The midplane density for a vertically isothermal disk is related to the surface density and vertical pressure scale height by
\begin{equation} \label{eq:rhomid}
    \rho = \dfrac{\Sigma}{\sqrt{2\pi} H},
\end{equation}
where the pressure scale height is $H = c_{s} / \Omega$. By utilizing Eq. (\ref{eq:toomreQ}) for $\Sigma$ in Eq. (\ref{eq:rhomid}), we obtain 
\begin{equation} \label{eq:dens}
    \rho = \dfrac{M_{*}}{\pi \sqrt{2\pi} Q} R^{-3},
\end{equation}
which yields an expression for the mass density at the disk outer edge 
\begin{equation} \label{eq:dens0}
    \rho_0 = \dfrac{M_{*}}{\pi \sqrt{2\pi} Q} R_0^{-3}.
\end{equation}
Substituting Eq.~(\ref{eq:dens0}) into Eq.~(\ref{eq:op1}), we obtain an expression for the dust opacity (cross-section per gram of gas and dust, not just dust) at a given wavelength 
\begin{equation}
    \kappa_\nu = 
        \dfrac{\pi \sqrt{2\pi} Q R_0^2}{M_{*}}\tau_{0,\nu} 
            = 1.7 \times 10^{-2} 
                Q 
                \tau_{0,\nu}
                \left(\dfrac{R_0}{70 \text{au}}\right)^2
                \left(\dfrac{0.25~M_\odot}{M_{*}}\right) 
                \dfrac{{\rm cm}^2}{\rm g}.
    \label{eq:op2}
\end{equation}
This expression forms the basis of our technique for inferring dust opacity from the characteristic optical depth $\tau_{0,\nu}$, which can be determined from the observed brightness temperature profile, as we illustrate with a simple semi-analytical model next\footnote{We should stress that this technique only yields a lower limit to the dust opacity unless the value of the Toomre parameter $Q\ (>1)$ is known or can be estimated (see Fig.~\ref{fig:opacities} below). In addition, our method of estimating the characteristic optical depth $\tau_{0,\nu}$ is specific to edge-on disks and should not be applied to non-edge-on disks without modifications.}. 
%
% At the risk of loss of generality...additional parameters...warning, generalized for lower mass edge-on disks.... edge-on....and mixing, latter.
%

\subsection{Model Setup}
\label{ssec:modelsetup}
 
To make the semi-analytical model as simple as possible, we will make a few simplifying assumptions. First, we will assume that the disk is viewed exactly edge-on, so that the problem becomes one dimensional (1D), with all physical quantities depending on the cylindrical radius $R$ on the disk midplane; the slight deviation (by $\sim 4^\circ$, \citealt{Lee2017_lane}) from being exactly edge-on is accounted for in the 2D model to be discussed in Section~\ref{sec:axisymmetric}. Second, we assume that the mass density is given by Eq.~(\ref{eq:dens}), with a spatially constant Toomre $Q$ parameter for the disk region probed by our multi-wavelength ALMA observations. This should be a reasonable approximation for relatively massive disks (relative to their central star) that are marginally gravitationally unstable, as appears to be the case for HH 212 mms \citep{Tobin2020}\footnote{As shown in Fig.~\ref{fig:semiana_fitmid} below, our multi-band ALMA continuum observations of HH 212 mms probe the outer disk region with a relatively limited range in radius, from $\sim 20$~au to $\sim 60$~au. The limited radial extent makes the constant $Q$ assumption more justifiable compared to the entire disk.}. It may break down for low-mass disks that are highly gravitationally stable, with $Q$ well above unity. The effects of a non-constant $Q$ are explored in Appendix \ref{sec:nonconstantQ}. 

%{dimensionless; what about B(T)?}

To cast the governing equations into a dimensionless form, we normalize all physical quantities by their values at the outer edge of the disk $R_0$. In the case of mass density, we have 
\begin{equation} \label{eq:rhoscaled}
    \rho = \rho_{0} \bigg( \dfrac{R}{R_{0}} \bigg)^{-3}.
\end{equation}
The distance from the center along the projected midplane is denoted by the impact parameter $b$, with $b=0$ corresponding to the center. The coordinate $z$ is the distance along a line of sight where the observer is located in $z = + \infty$ and $z=0$ corresponds to the line perpendicular to the line of sight that passes through the origin. For a given $b$, the radius at a location along the line of sight is $R = \sqrt{b^{2} + z^{2}}$.  We define a dimensionless distance $\hat{z} \equiv z / R_{0}$ and impact parameter $\hat{b} \equiv b / R_{0}$. 

The optical depth $\tau'_{\nu}(z)$ from a location on the disk at a distance $z$ along a line of sight with impact parameter $b$ to the observer is given by 
\begin{align} \label{eq:optdepth}
    \tau_{\nu}'(z) &= \int_{z}^{ \sqrt{R_{0}^{2} - b^{2}} } \rho \kappa_{\nu} dz \text{ .}
\end{align}
By applying Eq. (\ref{eq:rhoscaled}), we obtain the optical depth in dimensionless form:
\begin{align}
    \tau_{\nu}'(\hat{z}) = \dfrac{\tau_{0, \nu}}{\hat{b}^{2}} \bigg( \sqrt{1 - \hat{b}^{2}} - \dfrac{\hat{z}}{ \sqrt{\hat{b}^{2} + \hat{z}^{2}}} \bigg) \label{eq:optdepth_p3}
\end{align}
where $\tau_{0, \nu} \equiv \rho_{0} \kappa_{\nu} R_{0}$ is the characteristic optical depth defined earlier. It reflects the fact that density and opacity are degenerate and only their product (which is proportional to the optical depth) matters for the observed intensity.

To obtain the emergent intensity as a function of $\hat{b}$, we need to know the temperature, which controls the source function (taken as the Planck function). We adopt a power-law temperature profile 
\begin{equation}
    T = T_{0} \bigg( \dfrac{R}{R_{0}} \bigg)^{-q}
    \label{eq:temp}
\end{equation}
where $T_{0}$ is the temperature at the edge of the disk. The emergent intensity assuming black body radiation is 
\begin{equation} \label{eq:I_semi}
    I_{\nu} (\hat{b}) = \dfrac{2h \nu^{3}}{c^{2}}  
            \int_{-\sqrt{1 - \hat{b}^{2}}}^{\sqrt{1 - \hat{b}^{2}}}
            \dfrac{\tau_{0,\nu}}{(\hat{b}^{2} + \hat{z}^{2})^{3/2}}
            %\bigg[ \exp{\bigg[ x_{0} (\hat{b}^{2} + \hat{z}^{2})^{q/2} \bigg]} - 1 \bigg]^{-1} e^{-\tau'(\hat{z})} d\hat{z}
            \dfrac{ e^{-\tau'(\hat{z})} }{ \exp{\bigg[ x_{0} (\hat{b}^{2} + \hat{z}^{2})^{q/2} \bigg]} - 1 }
             d\hat{z}
\end{equation}
where $\nu$ is the observing frequency and $x_{0} \equiv h\nu / k T_{0} $. To develop insight, it is beneficial to first consider the Rayleigh-Jeans limit ($x_{0} << 1$), since the brightness temperature is linearly proportional to the intensity. In this limit, the brightness temperature of $I_{\nu}$ normalized by $T_{0}$ becomes
\begin{equation} \label{eq:tbsemi}
    \hat{T}_{\nu}(\hat{b}) \equiv \dfrac{\lambda^{2}}{2k T_{0}} I_{\nu} 
    = \int_{-\sqrt{1 - \hat{b}^{2}}}^{\sqrt{1 - \hat{b}^{2}}} \dfrac{\tau_{0,\nu}}{ (\hat{b}^{2} + \hat{z}^{2} )^{(q + 3)/2}}
        e^{
            - \tau'(\hat{z})
            }
                d \hat{z} .
\end{equation}
Obviously from Eq. (\ref{eq:tbsemi}), looking towards the edge of the disk, $\hat{b} \rightarrow 1$, the integration drops to zero. We see that the dimensionless brightness temperature distribution $\hat{T}_{\nu}(\hat{b})$ is uniquely determined by the characteristic optical depth $\tau_{0,\nu}$ for a given temperature power index $q$. We will take advantage of this important feature when we fit the multi-wavelength observations, which will allow us to constrain the outer radius of the disk $R_0$ and its temperature $T_0$, which are the scaling for the impact parameter and temperature, respectively.

The shape of the brightness temperature $\hat{T}_{\nu}(\hat{b})$ can be characterized by the impact parameter $\hat{b}_{1,\nu}$ along which line of sight the total optical depth becomes unity. It separates the the inner optically thick part of the disk where the brightness temperature is relatively constant from the outer optically thin part where it drops to zero towards the disk edge and thus provides a measure of the half-width of the brightness temperature profile. Using Eq. (\ref{eq:optdepth_p3}) and setting $\tau'_{\nu}(-\sqrt{1 - \hat{b}_{1}^{2}}) = 1$, we can solve for the characteristic impact parameter as:
\begin{equation} \label{eq:p3b1}
    \hat{b}_{1,\nu} = \sqrt{\dfrac{2}{1 + \sqrt{1 + 1/\tau_{0,\nu}^{2} }}} \text{ ,}
\end{equation}
in terms of the characteristic optical depth $\tau_{0,\nu}$. 
In the limit where $\tau_{0,\nu} \gg 1$, we have $\hat{b}_{1,\nu} \rightarrow 1$, which means the total optical depth $\tau$ is greater than $1$ along essentially all sight lines and the half-width of the profile is fixed with those $\tau_{0,\nu}$. In the opposite limit where the characteristic optical depth $\tau_{0,\nu} \ll 1$, the characteristic impact parameter becomes increasingly small, with $\hat{b}_{1,\nu} \rightarrow \sqrt{2 \tau_{0,\nu}}$. In other words, at longer wavelengths where $\tau_{0,\nu}$ decreases, the sight lines where $\tau > 1$ decreases and thus the half-width of the profile also decreases. In fact, since $\hat{b}_{1}$ is determined by $\tau_{0,\nu}$ which in turn depends on $\kappa_{\nu}$, the frequency dependence of $\hat{b}_{1}$ is directly related to the opacity index $\beta$. The frequency dependence can be determined by 
\begin{equation}
    \dfrac{ \partial \ln \hat{b}_{1} }{ \partial \ln \nu } 
        = \dfrac{\beta}{2} \dfrac{1}{ \tau_{0,\nu} \sqrt{ \tau_{0,\nu}^{2} + 1 } + \tau_{0,\nu}^{2} + 1 } \text{ .}
\end{equation}
When $\tau_{0,\nu} \gg 1$, $\partial \ln \hat{b}_{1} / \partial \ln \nu \rightarrow 0$ which means the width of the profile does not change with frequency and stays at $\hat{b}_{1}=1$ from Eq.~(\ref{eq:p3b1}). In the other limit, when $\tau_{0,\nu} \ll 1$, $\partial \ln \hat{b}_{1} / \partial \ln \nu \rightarrow \beta / 2$, i.e., how quickly the width of the profile decreases towards longer wavelength is half of how fast opacity decreases. 

In the optically thick region interior to $\hat{b}_{1,\nu}$, we can use the Eddington-Barbier relation to estimate the brightness temperature from the temperature at the $\tau'=1$ surface, which we denote by its dimensionless distance along the line of sight $\hat{z}_{1,\nu}$. An analytical expression of $\hat{z}_{1,\nu}$ can also be determined from Eq. (\ref{eq:optdepth_p3}) at an impact parameter of $\hat{b}$:
\begin{align}
    \hat{z}_{1,\nu} = \dfrac{
                 \sqrt{1 - \hat{b}^{2}} - \dfrac{\hat{b}^{2}}{\tau_{0,\nu}}
            }{
                \sqrt{ 1 - \bigg(\dfrac{\hat{b}}{ \tau_{0,\nu} } \bigg)^{2} + \dfrac{2}{ \tau_{0,\nu} } \sqrt{1 - \hat{b}^{2}} }
                \text{ .} }
\end{align}
When $\tau_{0,\nu} \gg 1$, $\hat{z}_{1,\nu}$ approaches $\sqrt{1-\hat{b}^{2}}$, i.e., the $\tau'=1$ surface traces the very edge of the circular disk. 

For a given $\tau_{0,\nu}$, the $\tau'=1$ surface along the line of sight towards the center of the disk is located at a radius 
\begin{equation}
    \hat{R}_{1,\nu} \equiv \hat{z}_{1,\nu}(\hat{b} = 0)
        = \frac{1}{\sqrt{ \dfrac{2}{\tau_{0,\nu} } + 1}} 
        = \frac{1}{\sqrt{2\sqrt{ \bigg( \dfrac{2}{ \hat{b}_{1,\nu}^{2} } -1 \bigg)^2-1} + 1 }}
    \label{eq:rbrelation}
\end{equation}
where the last equality was obtained by replacing $\tau_{0,\nu}$ with $\hat{b}_{1,\nu}$ using Eq.~(\ref{eq:p3b1}). This relation is particularly simple in two limits: $\hat{R}_{1,\nu}\rightarrow 1$ as $\hat{b}_{1,\nu}\rightarrow 1$ in the limit $\tau_{0,\nu}\gg 1$ (which makes sense because the $\tau'=1$ surface is located near the outer edge), and $\hat{R}_{1,\nu} \rightarrow \frac{\hat{b}_{1,\nu}}{2}$ in the limit $\tau_{0,\nu}\ll 1$, which means that the $\tau'=1$ surface along the line of sight to the center is located at a radius that is half of the characteristic half-width of the brightness temperature profile along the disk major axis (see Fig.~\ref{fig:semiana_opt_normmax}b below for an illustration). 

The relation between $\hat{R}_{1,\nu}$ and $\hat{b}_{1,\nu}$ from Eq. (\ref{eq:rbrelation}) illustrates an important conceptual point for edge-on disks --- the readily observable brightness temperature distribution along the major axis in the plane of the sky (see Fig.~\ref{fig:tbcut}a) enables a determination of the radius $R_{1,\nu}$ of the $\tau'=1$ surface along the line of sight to the center, where the temperature can be estimated from the observed brightness temperature along the same sight line.

With well-resolved observations at multiple wavelengths that have different opacities, we should in principle be able to determine the temperatures at different $\tau'=1$ surfaces at different radii and thus the radial distribution of the temperature along the disk midplane. We derive an analytical expression for the brightness temperature towards the center in Appendix \ref{sec:qobs} and point out the main results here. The peak brightness temperature is  
\begin{align} \label{eq:Thatb0p3}
    \hat{T}_{\nu}(\hat{b}=0)
        &=  \bigg(\dfrac{\tau_{0,\nu}}{2} \bigg)^{ -\dfrac{q}{2} } e^{ \dfrac{\tau_{0,\nu} }{2} }
            \Gamma \bigg( \dfrac{q}{2}+1, \dfrac{ \tau_{0,\nu} }{2} \bigg)
\end{align}
where $\Gamma(a,x) = \int_{x}^{\infty} z^{a-1} e^{-z} dz $ is the upper incomplete gamma function. For a finite $q$, when $\tau_{0,\nu} \gg 1$, we have $\hat{T}_{\nu}(\hat{b}=0) \sim 1 + q/\tau_{0,\nu}$ which means the observed brightness temperature is simply the temperature at the edge of the disk $T_{0}$. Since $\hat{R_{1}}$ and $\hat{T}_\nu$ can be determined at $\hat{b}=0$ for different wavelengths, one can calculate an observed temperature power-law index defined by 
\begin{equation} \label{eq:qobsdef}
    q_{\text{obs}} \equiv - \dfrac{ \partial \ln \hat{T}_{\nu} |_{\hat{b}=0} }{ \partial \ln \hat{R}_{1} } 
        %= - \dfrac{ \partial \ln \hat{T_{\nu}} |_{\hat{b}=0} / \partial \tau_{0,\nu} }{ \partial \ln \hat{R_{1}} / \partial \tau_{0,\nu} }
\end{equation}
which yields $q_{\text{obs}} \sim q$ (see Appendix \ref{sec:qobs} for details). This demonstrates why $q$ can be measured for edge-on disks with multiple wavelength observations. 
%can also be extended to any value of $p$. For our considered case of $p=3$, we have 

\subsection{Illustration of model features} 
\label{ssec:illustration}

\begin{figure*}
    \centering
    \includegraphics[width=\textwidth]{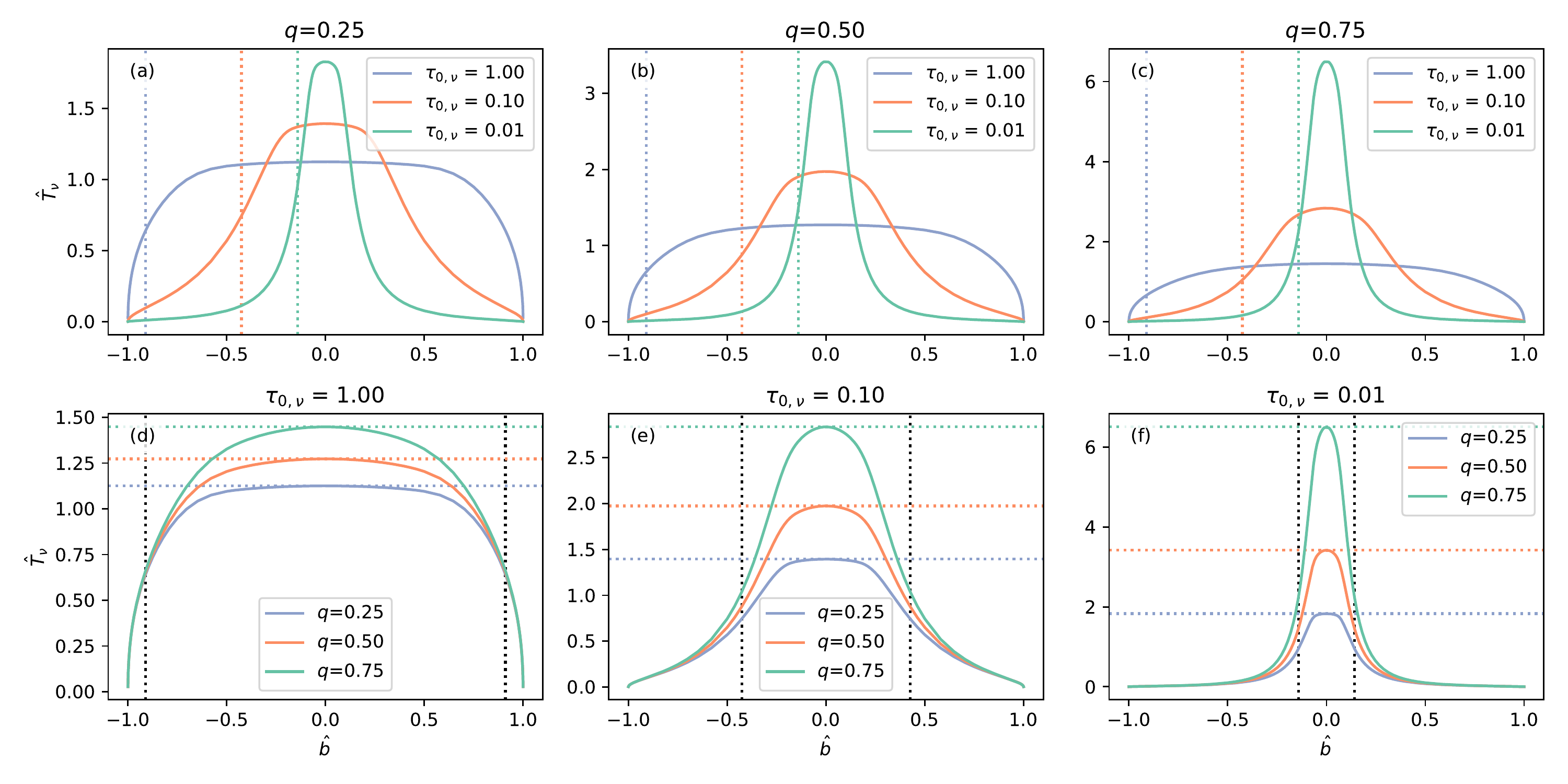}
    \caption{
        The normalized brightness temperature $\hat{T}_{\nu}$ as a function of the normalized impact parameter $\hat{b}$ along the major axis of an edge-on disk from the semi-analytical model. The top row compares the different values of $\tau_{0,\nu}$ for a given $q$. The vertical dotted lines mark where the optical depth becomes 1 for each $\tau_{0,\nu}$ based on Eq. (\ref{eq:p3b1}), or $\hat{b}_{1}$. The bottom row compares the different values of $q$ for a given $\tau_{0,\nu}$. The black vertical dotted lines are $\hat{b}_{1}$ for each $\tau_{0,\nu}$. The horizontal dotted lines are the brightness temperature at $\hat{b}=0$ obtained analytically using Eq. (\ref{eq:Thatb0p3}) and are color coded to the corresponding $q$ in each panel. 
            }
    \label{fig:semiana_num}
\end{figure*}

\begin{figure}
    \centering
    \includegraphics[width=0.9\columnwidth]{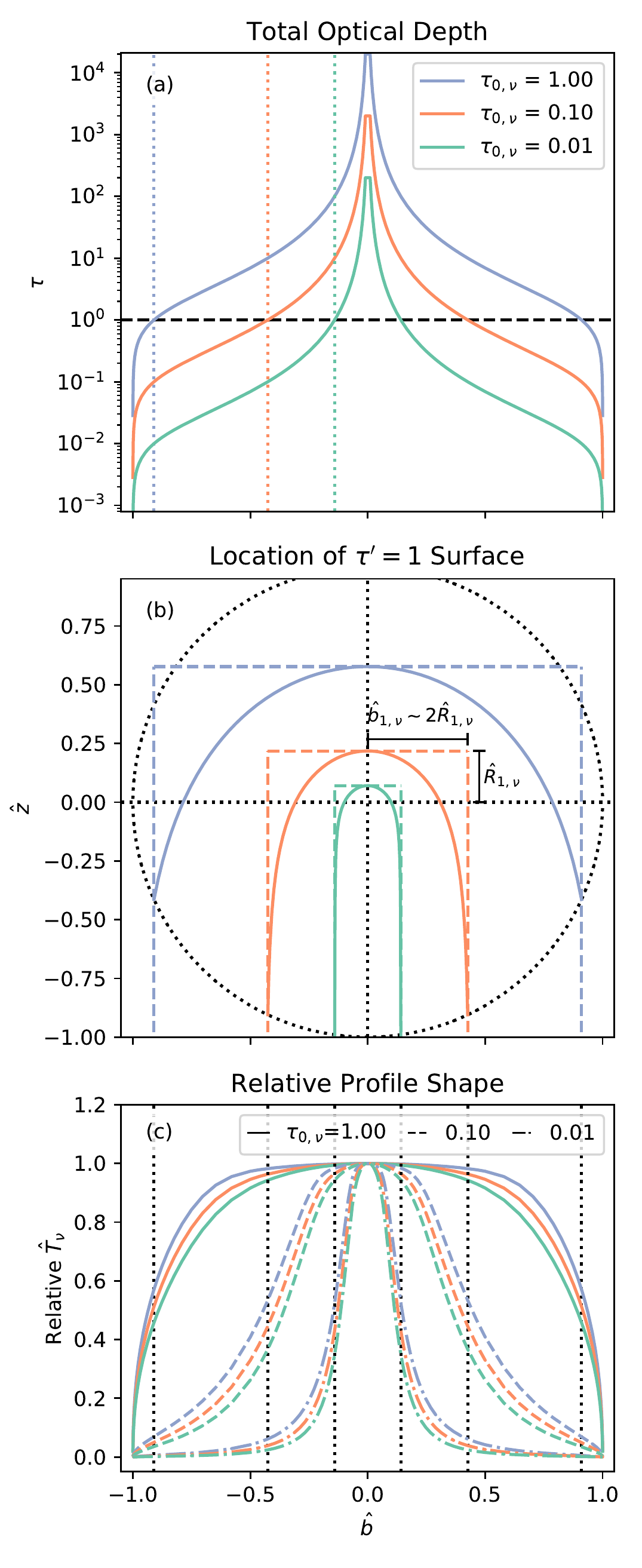}
    \caption{
        Panel (a) shows the total optical depth $\tau$ along the normalized impact parameter $\hat{b}$ for different values of $\tau_{0,\nu}$. The horizontal dashed line marks where $\tau=1$. The vertical dotted lines are $\hat{b_{1}}$ color coded to the corresponding $\tau_{0,\nu}$. Panel (b) shows the location of the $\tau'=1$ surfaces for different values of $\tau_{0,\nu}$ as viewed from $\hat{z} \rightarrow \infty$. The dotted circle denotes the edge of the disk. The horizontal dotted line is where $\hat{z}=0$ and the vertical dotted line is where $\hat{b}=0$. The vertical dashed lines are $\pm \hat{b}_{1}$ color coded to the corresponding $\tau_{0,\nu}$. The corresponding horizontal dashed lines are $\hat{R_{1}}$. We specifically mark the $\hat{b}_{1,\nu}$ and $\hat{R}_{1,\nu}$ for $\tau_{0,\nu}=0.1$ as an example. Recall that $\hat{b}_{1,\nu} \rightarrow 2 \hat{R}_{1,\nu}$ when $\tau_{0,\nu} << 1$ based on Eq. (\ref{eq:rbrelation}). 
        Panel (c): The normalized brightness temperature $\hat{T}_{\nu}$ relative to its own peak for the same combinations of $\tau_{0,\nu}$ and $q$ in Fig. \ref{fig:semiana_num}. The profiles with $\tau_{0,\nu}=$ 1, 0.1, and 0.01 are plotted in solid, dashed, and dashed-dotted lines. The blue, orange, and green lines correspond to $q=$ 0.25, 0.5, and 0.75. The black dotted vertical lines are $\hat{b}_{1}$ for each $\tau_{0,\nu}$.  
            }
    \label{fig:semiana_opt_normmax}
\end{figure}

The semi-analytical model for the brightness temperature distribution along the major axis of an edge-on disk is controlled by two free parameters, the characteristic optical depth at a given frequency $\tau_{0,\nu}$ and the power-law index of the radial temperature distribution $q$. The latter can in principle be determined from the observational data, but for illustration purposes in this subsection, we will leave it as a free parameter and consider three representative values $q = $ 0.25, 0.5, and 0.75. The latter two represent the expected power-law indices for a passive disk irradiated by the stellar radiation ($q=0.5$) and an active disk heated mainly by accretion ($q=0.75$; \citealt{Armitage2015}). In the top row of Fig. \ref{fig:semiana_num}, we compare the different brightness temperature profiles with three different characteristic optical depths $\tau_{0,\nu}=0.01, 0.1$ and 1 for each $q$. Starting with Fig. \ref{fig:semiana_num}a, the brightness temperature profile in the most optically thick ($\tau_{0,\nu}=1$) and most flat radial temperature distribution ($q=0.25$) of the three cases has a broad plateau where $\hat{T}_\nu$ is roughly constant. The main reason for this feature is that the impact parameter where the total optical depth through the (edge-on) disk becomes unity, $\hat{b}_{1}$, is close to 1, and thus most of the disk along the equator is optically thick. This is illustrated more quantitatively in Fig. \ref{fig:semiana_opt_normmax}a, where the total optical depth is plotted as a function of the impact parameter (see the blue curve). The plateau occurs because over most of the $\tau'=1$ surface (shown as the blue curve in Fig.~\ref{fig:semiana_num}b) there is relatively little variation in distance from the center, which corresponds to a relatively small variation in temperature, especially for a relatively flat radial temperature distribution. The combination of a large optical depth over most sight lines and relatively flat temperature distribution explains the ``boxiness" of the profile. In the limit where $\tau_{0,\nu} \rightarrow \infty$, we have $\hat{b}_{1}\rightarrow 1$, with the profile approaching an actual box function with a sharp edge. A boxy profile, if observed, would provide a strong indication of a mostly optically thick disk with a well-defined outer 
edge. 

%\begin{figure}
%    \centering
%    \includegraphics[width=\columnwidth]{figures/semiana_opt.png}
%    \caption{
%        Panel (a) shows the total optical depth $\tau$ along the normalized impact parameter $\hat{b}$ for different values of $\tau_{0,\nu}$. The horizontal dashed line marks where $\tau=1$. The vertical dotted lines are $\hat{b_{1}}$. Panel (b) shows the location of the $\tau'=1$ surfaces for different values of $\tau_{0,\nu}$ as viewed from $\hat{z} \rightarrow \infty$. The dotted circle denotes the edge of the disk. The horizontal dotted line is where $\hat{z}=0$ and the vertical dotted line is where $\hat{b}=0$. The vertical dashed lines are $\pm \hat{b}_{1}$ color coded to the corresponding $\tau_{0,\nu}$. The corresponding horizontal dashed lines are $\hat{R_{1}}$. We specifically mark the $\hat{b}_{1,\nu}$ and $\hat{R}_{1,\nu}$ for $\tau_{0,\nu}=0.1$ as an example. Recall that $\hat{b}_{1,\nu} \rightarrow 2 \hat{R}_{1,\nu}$ when $\tau_{0,\nu} << 1$ based on Eq. (\ref{eq:rbrelation}). 
%            }
%    \label{fig:semiana_opt}
%\end{figure}

As the characteristic optical depth $\tau_{0,\nu}$ decreases, which is true at a longer wavelength where the opacity is smaller, the characteristic half-width $\hat{b}_{1}$ decreases and the outer regions of the disk become more optically thin (compare the yellow and green curves with the blue curve in the upper panels of Fig.~\ref{fig:semiana_num} and in both panels of Fig.~\ref{fig:semiana_opt_normmax}). The optically thin regions do not produce much $\hat{T}_{\nu}$, which is part of the reason for the decrease in the width of the profile. Another part is that, as $\tau_{0, \nu}$ decreases, the $\tau'=1$ surface moves closer to the center (see Fig.~\ref{fig:semiana_opt_normmax}b), where the temperature is higher, which leads to a higher peak brightness temperature along the line of sight to the center. 

%\begin{figure}
%    \centering
%    \includegraphics[width=\columnwidth]{figures/semiana_normmax.png}
%    \caption{
%        The normalized brightness temperature $\hat{T}_{\nu}$ relative to its own peak for the same combinations of $\tau_{0,\nu}$ and $q$ in Fig. \ref{fig:semiana_num}. The profiles with $\tau_{0,\nu}=$ 1, 0.1, and 0.01 are plotted in solid, dashed, and dashed-dotted lines. The blue, orange, and green lines correspond to $q=$ 0.25, 0.5, and 0.75. The black dotted vertical lines are $\hat{b}_{1}$ for each $\tau_{0,\nu}$.  
%            }
%    \label{fig:semiana_normmax}
%\end{figure}

Comparing across Fig. \ref{fig:semiana_num}a, b, and c, the qualitative behaviors are all the same regardless of the value of the temperature power-law index $q$. The difference caused by the power-law index is in the contrast of the $\hat{T}_\nu$ peak between different values of $\tau_{0,\nu}$. For $q=0.25$, the peak $\hat{T}_\nu$ of the $\tau_{0,\nu}=0.01$ profile is $\sim 1.5$ times the peak $\hat{T}_\nu$ of the $\tau_{\nu, 0}=1$ profile (see Fig.~\ref{fig:semiana_num}a), whereas ratio of the same pair for $q=0.75$ is $\sim 4$ times (see Fig.~\ref{fig:semiana_num}c). We can understand this by looking at Fig. \ref{fig:semiana_num}d, e, and f. For a given $\tau_{0, \nu}$, the total optical depth is the same and $\hat{z}_{1,\nu}$ is also the same. With $\tau_{0, \nu} = 0.01$ (Fig. \ref{fig:semiana_num}d), the $q=0.75$ profile has the highest peak $\hat{T}_\nu$ simply because the temperature increased more rapidly to $\hat{z}_{1,\nu}$ from $T_{0}$ at the edge (Eq.~\ref{eq:temp}). For $\tau_{0, \nu} = 1$ (Fig. \ref{fig:semiana_num}f), the difference is less sensitive to $q$ because the $\tau'=1$ surface is closer to the outer edge (see Fig.~\ref{fig:semiana_opt_normmax}b), which reduces the sensitivity of the peak brightness temperature to the index $q$.

\subsection{Application to ALMA Band 7, 6 and 3 data}
\label{ssec:application}

\begin{figure}
    \centering
    \includegraphics[width=\columnwidth]{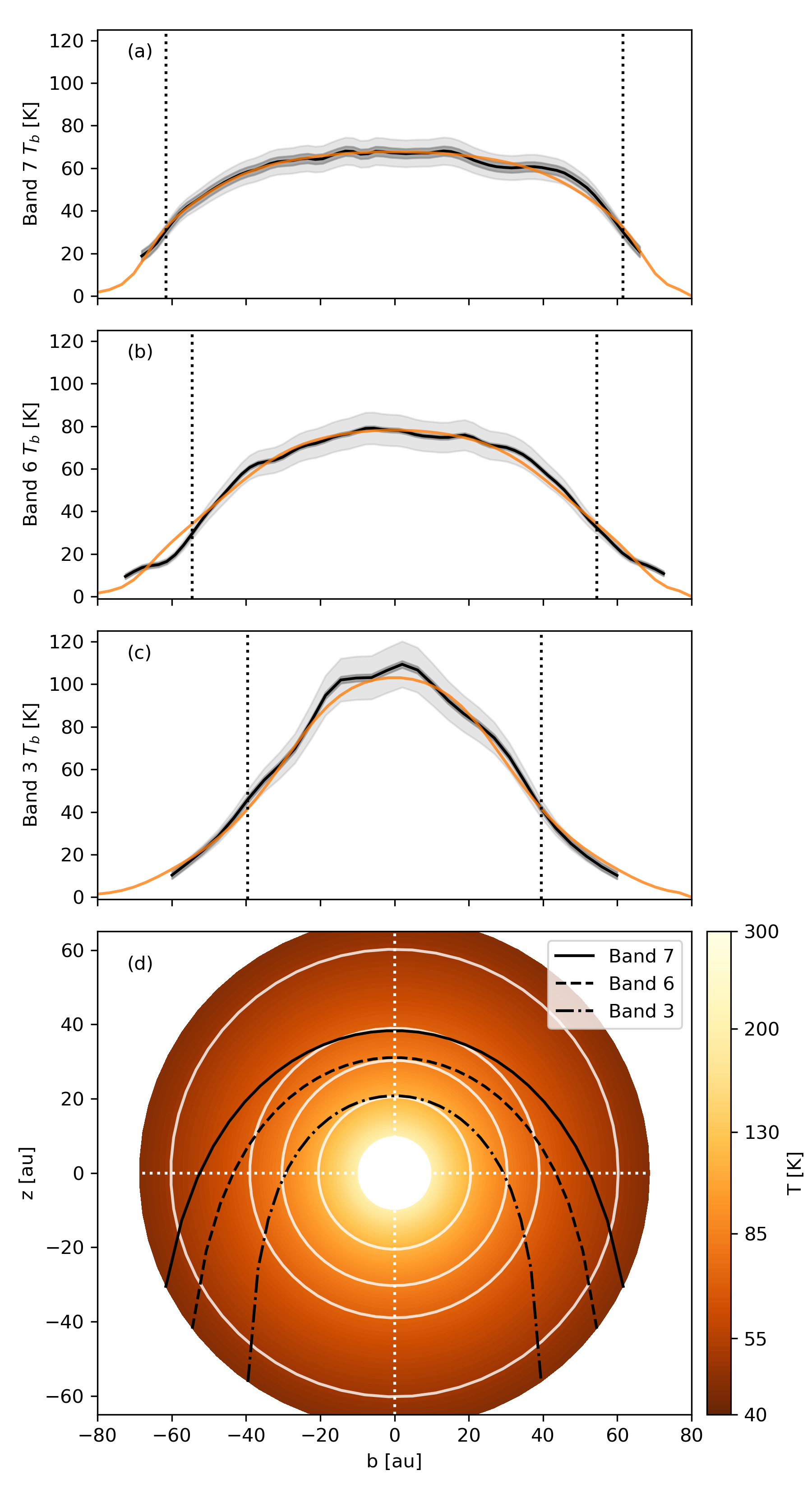}
    \caption{
        The brightness temperature using the Planck function along the disk major axis comparing the semi-analytical model and Bands 7 (panel a), 6 (panel b), and 3 (panel c) observations. The black solid line is the observations and the shaded regions are the adopted uncertainties for the noise (darker shade) and absolute flux calibration (lighter shade) respectively. The orange line is the model result. The vertical dotted lines are $\pm \hat{b}_{1,\nu}$ determined from Eq. (\ref{eq:p3b1}) scaled to the fitted $R_{0}$.  
        Panel d: The midplane temperature structure derived from fitting the major axis cuts using the 1D semi-analytical model. The color scale is the temperature in Kelvin. The white contours show the temperature at 50, 70, 80, and 115 Kelvin. The $\tau'=1$ surface traced by Bands 7, 6, and 3 (as viewed by an observer located at $z=\infty$) are plotted in black solid, dashed, and dash-dotted lines.
            }
    \label{fig:semiana_fitmid}
\end{figure}

It is immediately clear from Fig.~\ref{fig:semiana_num} that the above semi-analytical model qualitatively captures the two main features of the brightness temperature profiles observed along the major axis of the HH 212 disk at different wavelengths (shown in Fig.~\ref{fig:tbcut}a), namely, the profile becomes more centrally peaked and the peak value increases as the wavelength increases and the corresponding optical depth drops (see, e.g., Fig.~\ref{fig:semiana_num}b). Qualitatively, it is straightforward to see why one can constrain the two dimensionless quantities -- the temperature power-index $q$ and the characteristic optical depth $\tau_{0,\nu}$ at that wavelength -- from the shape of the observed profile at a given wavelength, e.g., ALMA Band 7 (shown in Fig.~\ref{fig:semiana_fitmid}a below). For example, the Band 7 profile is not as peaky as the $\tau_{0,\nu}=0.1$ profiles shown in Fig.~\ref{fig:semiana_num}e and more similar to the $\tau_{0,\nu}=1.0$ profiles shown in Fig.~\ref{fig:semiana_num}f, indicating that the value of its $\tau_{0,\nu}$ must be somewhere around 1. In Fig. \ref{fig:semiana_opt_normmax}c, we plot the profiles relative to its own peak to examine only the shape of the profile. Both $\tau_{0,\nu}$ and $q$ affect the shape of a profile, but $\tau_{0,\nu}$ plays the larger role in determining the shape of the profile given that it can vary by an order of magnitude due to opacity. The index $q$ mainly controls the relative peak values of the brightness temperature at different wavelengths (that have different values of $\tau_{0,\nu}$), with a larger $q$ (or a steeper radial temperature gradient) yielding a larger difference in peak brightness temperature between two wavelengths (compare, e.g., the red and blue curves in the top panels of Fig.~\ref{fig:semiana_num}). It can be determined from the ratio of the peak brightness temperatures at two wavelengths, such as Band 7 and 6 (shown in Fig.~\ref{fig:semiana_fitmid}b below), after the difference in $\tau_{0,\nu}$ between the two is taken into account. Once $\tau_{0,\nu}$ and $q$ are determined, it is straightforward to obtain the radius $R_0$ and temperature $T_0$ at the disk outer edge, which sets the scaling for the length and brightness temperature, respectively.

%\begin{figure*}
%    \centering
%    \includegraphics[width=\textwidth]{figures/semiana_fit.png}
%    \caption{
%        The brightness temperature using the Planck function along the disk major axis comparing the semi-analytical model and Bands 7 (panel a), 6 (panel b), and 3 (panel c) observations. The black solid line is the observations and the shaded regions are the adopted uncertainties for the noise (darker shade) and absolute flux calibration (lighter shade) respectively. The orange line is the model result. The vertical dotted lines are $\pm \hat{b}_{1,\nu}$ determined from Eq. (\ref{eq:p3b1}) scaled to the fitted $R_{0}$.  
%            }
%    \label{fig:semiana_fit}
%\end{figure*}

As a quantitative example, we apply the semi-analytic model to fit the flux density profiles at Bands 7, 6, and 3. The first two images are better spatially resolved than other Bands, while Band 3 suffers from finite resolution which we will address in a more complete way in Section \ref{sec:axisymmetric}. We use the \texttt{optimize.least\_squares} from the \texttt{scipy} \texttt{python} package to find the best fit parameters. To estimate the uncertainty of the fitted parameters, we consider only the noise uncertainty and sample the major axis in steps of a beam size to avoid correlated noise in the image plane. One should note that comparisons across different observations require an additional absolute flux uncertainty. We do not include it in the uncertainty estimates, since it simply scales the image and does not affect the relative points along an image profile. As described above, since the shape of the profile is determined mainly by $\tau_{0,\nu}$ and the peak of the profile is scaled by $T_{0}$, the absolute flux uncertainty mainly affects $T_{0}$ and $q$ and not $\tau_{0,\nu}$. In Fig.~\ref{fig:semiana_fitmid}a-c, we show the noise uncertainty by a dark shaded region and a lighter shaded region to show the uncertainty including a $10\%$ absolute flux uncertainty added in quadrature with the noise. We ignore regions with emission less than 5 times the noise uncertainty. To produce the model fits, we calculate Eq. (\ref{eq:I_semi}) and convolve the profile with a Gaussian beam to include effects of finite resolution in 1D. A distance of 400 pc is assumed. The fitted values and its uncertainty are listed in Table \ref{tab:curvefit}. 

\begin{table}
    \centering
    \begin{tabular}{l c c}
        Parameter & Variable & Best fit value\\
        \hline
        Disk Outer Radius [au] & $R_{0}$ & 68.7 $\pm$ 0.6\\
        Temperature at $R_{0}$ [K] & $T_{0}$ & 45.2 $\pm$ 0.7\\
        Temperature Power-law Index & $q$ & 0.77 $\pm$ 0.02 \\
        Characteristic Optical Depths & $\tau_{0,\text{B7}}$ & 0.90 $\pm$ 0.06 \\
            & $\tau_{0,\text{B6}}$ & 0.52 $\pm$ 0.02 \\
            & $\tau_{0,\text{B3}}$ & 0.20 $\pm$ 0.01
    \end{tabular}
    \caption{
        The best fit values for Bands 7, 6, and 3 using the semi-analytical model. The uncertainty is estimated by considering noise. 
            }
    \label{tab:curvefit}
\end{table}

Fig. \ref{fig:semiana_fitmid}a-c show the fitted profiles compared to the Bands 7, 6 and 3 observations in brightness temperature using the Planck function. We see that the model can reproduce the key characteristics, including the peak, width, and shape, fairly well. 
%Some deviations occur towards the wings of the profiles. In particular, where the model drops to zero at a finite impact parameter $b$, the observed profile has some remaining excess. This is likely due to contributions from the envelope that is not included in the model.

To further illustrate the relation between the profile and the temperature, we show the resulting temperature structure in the midplane and the $\tau'=1$ surface traced at each wavelength in Fig.~\ref{fig:semiana_fitmid}d. One can clearly see that the $\tau'=1$ surfaces near $b=0$ are nearly circular which produces the signature flat plateau. Furthermore, the region within $\sim 20$ au of the star remains un-probed because the disk is optically thick even at Band 3. 

%\begin{figure}
%    \centering
%    \includegraphics[width=\columnwidth]{figures/semiana_mid.png}
%    \caption{
%        The midplane temperature structure derived from fitting the major axis cuts using the 1D semi-analytical model. The color scale is the temperature in Kelvin. The white contours show the temperature at 50, 70, 80, and 115 Kelvin. The $\tau'=1$ surface traced by Bands 7, 6, and 3 (as viewed by an observer located at $z=\infty$) are plotted in black solid, dashed, and dash-dotted lines.
%            }
%    \label{fig:semiana_mid}
%\end{figure}

%\red{What are the numbers and why that make sense?}

The best fit parameters in Table~\ref{tab:curvefit} make physical sense. For example, the best fit characteristic optical depth for Band 7 is $\tau_{0, \text{B7}}\approx 0.9$, which indeed lies around 1, as anticipated above based on the shape of its brightness temperature profile. Since Band 6 and Band 3 have more peaky brightness temperature profiles, it should have a smaller $\tau_{0,\nu}$, which is indeed the case ($\tau_{0, \text{B6}}\approx 0.52$ and $\tau_{0,\text{B3}} \approx0.20$). The corresponding $\tau'=1$ surfaces are located, respectively, at radii $R_{1,\text{B7}} \sim 0.56 R_0$, $R_{1,\text{B6}} \sim 0.45 R_0$, and $R_{1,\text{B3}} \sim 0.30 R_{0}$ along the line of sight to the center, where the dust temperature should be close to the observed peak brightness temperature of $T_{b,B7}=67$K, $T_{b,B6}=78$K, and $T_{b,B3}=103$K, respectively. The expected temperature power-law index is then $ (\ln T_{b,\nu}-\ln T_{b,\nu})/(\ln R_{1,\nu}-\ln R_{1,\nu}) \sim 0.7$, which is indeed close to the best fit value of $\sim 0.77$. Extrapolating the power-law to the disk outer edge $R_0$ yields an expected temperature of $\sim 40$K, which is close to the best fit value of $45$K. In addition, the model fits yield a well-defined disk radius of $0.17$ arcsec, corresponding to $\sim 69$~au for an adopted distance of 400~pc.

With the disk radius $R_0$ and the characteristic optical depth $\tau_{0,\nu}$ at a given wavelength determined, one can put a constraint on the dust opacity $\kappa_\nu$ using Eq.~(\ref{eq:op2}). From the best fit values, we infer a dust opacity of $\kappa_\nu \sim 1.5\times 10^{-2} Q(0.25~M_\odot/M_{*})$~cm$^2$/g for Band 7 ($\lambda=0.85$~mm), $8.7\times 10^{-3} Q(0.25~M_\odot/M_{*})$~cm$^2$/g for Band 6 (1.3~mm), and $3.4\times 10^{-3} Q(0.25~M_\odot/M_{*})$~cm$^2$/g for Band 3 (3~mm). The constraints on dust opacities will be discussed in more detail in Section~\ref{ssec:opacity} below. 

We should note that the observed dust emission at a given wavelength mostly comes from the outer part of the disk that is not too optically thick along the line of sight for that wavelength. For example, in the model shown in Fig.~\ref{fig:semiana_fitmid}d, Band 7 probes the dust emission from the disk outer edge (at $\sim 70$~au) down to a radius of $\sim 40$~au. If there is a spatial variation of the dust opacity at Band 7 (arising, e.g., from a possible radial stratification of the grain size; \citealt{Carrasco2019}), the inferred opacity should be viewed as an ``average" beyond $\sim 40$~au. Similarly, the opacity inferred at Band 3 should be viewed as an ``average" beyond a radius of $\sim 20$~au.

% ================================================
\section{2D Modeling of Multi-wavelength Dust Emission} \label{sec:axisymmetric}
% include band 9, 3, and band Ka

The one-dimensional (1D) semi-analytical model in the last section provided a simple way to understand the major features of the brightness temperature distribution along the major axis and to use the profiles to infer the disk properties, including the existence and radius of an outer edge, a temperature gradient, and dust opacities at different wavelengths. There are, however, two limitations to this approach. First, it assumed that the disk is perfectly edge-on, with the line of sight lying on the disk midplane. While this is a reasonable first approximation, it is not strictly valid because the disk inclination angle is estimated to be $i\approx 86^{\circ}$ \citep{Lee2017_lane}. The deviation from exact edge-on orientation is crucial for explaining the pronounced asymmetry in brightness observed in Band 7 and 6, and to a lesser extent, Band 3 above and below the equatorial plane (see Fig.~\ref{fig:realImages}). Second, it assumed that the disk is well resolved vertically, which is likely appropriate for Band 7 and 6, but less so for other Bands where the lower resolution can lead to blending of the dust emission from the hotter surface layers with that from the cooler midplane. 

To account for both the disk inclination and the smearing effects of the finite telescope beam, we need to prescribe the vertical structure of the disk in addition to the midplane structure first described in Section \ref{sec:semianalytical}.  We will perform two dimensional (2D) radiative transfer modeling of the multi-wavelength dust emission using \textsc{RADMC-3D}\footnote{\textsc{RADMC-3D} is a publicly available code for radiative transfer calculations: \url{http://www.ita.uni-heidelberg.de/~dullemond/software/radmc-3d/}}. The modeling of both the major axis and minor axis cuts provides a rare opportunity to constrain the disk vertical temperature gradient. 

\subsection{Disk Model}

We assume that the disk is axisymmetric and adopt a cylindrical coordinate system $(R, z)$. We assume the vertical distribution of gas is in hydrostatic equilibrium, so that 
\begin{equation} \label{eq:hydro}
    \dfrac{\partial P}{\partial z} = - \rho \dfrac{G M_{*}}{R^{2} + z^{2}} \dfrac{z}{\sqrt{R^{2} + z^{2}}}
\end{equation}
where $P$ is the pressure related to density by $P = \rho c_{s}^{2}$. The isothermal sound speed, $c_{s}$, is determined by the temperature structure
\begin{equation}
    c_{s} = \sqrt{\dfrac{k T}{\mu m_{p}}}
\end{equation}
where $\mu m_{p}$ is the mean molecular weight. Eq. (\ref{eq:hydro}) can be reorganized to explicitly calculate the vertical variation of density 
\begin{equation}
    \dfrac{\partial \ln \rho}{\partial z} = - \dfrac{\partial \ln T}{\partial z} - \dfrac{1}{c_{s}^{2}} \dfrac{G M_{*} z}{(R^{2} + z^{2})^{3/2}} \text{ ,}
\end{equation}
while the midplane density is prescribed as Eq. (\ref{eq:rhoscaled}) as in the semi-analytical model. We assume that the dust is well-mixed with the gas and thus the vertical dust distribution is obtained.

For the temperature distribution, we will adopt simple parameterization and use the multi-wavelength dust continuum observations to constrain the free parameters in the distribution. We adopt a profile for the midplane temperature, $T_{\text{mid}}$ and another profile for the temperature in the atmosphere near the disk surface, $T_{\text{atm}}$:
\begin{align} \label{eq:temp2d_Tmid}
    T_{\text{mid}}(R) = T_{0, \text{mid}} \bigg( \dfrac{R}{R_{0}} \bigg)^{-q_{\text{mid}}} \\
    T_{\text{atm}}(r) = T_{0, \text{atm}} \bigg( \dfrac{r}{R_{0}} \bigg)^{-q_{\text{atm}}}
        \label{eq:temp2d_Tatm}
\end{align}
where $r = \sqrt{R^{2} + z^{2}}$ is the spherical radius. The temperature transitions from the midplane temperature to the atmospheric temperature by
\begin{equation} \label{eq:temp2d}
    T(R,z) = 
    \Bigg\{
        \begin{array}{l l l}
            T_{\text{atm}} + (T_{\text{mid}} - T_{\text{atm}}) \cos^{2\delta} \bigg( \dfrac{\pi z}{2 R h_{s} } \bigg) &\text{if } &  z/R < h_{s} \\
            T_{\text{atm}} & \text{if } & z/R \ge h_{s}
        \end{array}
\end{equation}
where $h_{s}$ is the dimensionless height where the temperature becomes fully described by $T_{\text{atm}}$ and $\delta$ is the sharpness of the transition, i.e., a small $\delta$ means a gradual transition and vice versa \citep[e.g.][]{Dartois2003, Rosenfeld2013}. Since $\delta$ and $h_{s}$ are correlated, we fix $\delta=2$ and for simplicity we keep $h_{s}$ radially independent to limit the number of free parameters. 

The dust opacity is determined from the characteristic optical depth $\tau_{0,\nu}\equiv \kappa_\nu\rho_0 R_0$ defined in Section~\ref{sec:semianalytical}, which is a free parameter for each of the wavelengths. We fixed the inclination to be $i=86^{\circ}$ and the stellar mass as $M_*=0.25 M_{\odot}$. 
After producing the model images, we convolve them with matching FWHM Gaussian beams from the observations. We fit Bands 7, 6, and 3 only as in Section \ref{sec:semianalytical}. We ignored Band 9, because it is contaminated by the envelope emission (that is not included in the model) due to the large telescope beam and high dust opacity at the (shortest) wavelength. We chose not to include Band Ka, because the free-free contamination could be substantial (up to $20\%$ of the total flux based on the empirical estimate in \citealt{Tychoniec2018}). To accurately incorporate the thermal emission from Band Ka would require high-resolution multi-wavelength observations at longer wavelengths to remove the free-free emission\footnote{Note that the observed peak brightness temperature is lower than that at Band 3 in Fig. \ref{fig:tbcut} even though Band Ka should be probing the inner regions of the disk closer to the central star where the temperature is higher because of its lower opacity and should have a larger contribution from free-free emission because of its longer wavelength. The reason for this apparent discrepancy is that most of the region within the VLA beam is optically thin with low emission, especially along the minor axis, which leads to a large reduction of the peak brightness temperature. It can be further reduced by scattering if the grains in the region probed by the Band Ka have grown to cm sizes.}, which we hope to do in the near future through an approved VLA program.

\subsection{Results} \label{ssec:2dresults}

%\red{1. Discuss how the best fit model is chosen; 2. connection between 1D and 2D models: How different is the 1D midplane compared to LOS in 2D models? What is the effect of adding Band Ka? 3. what is the conclusion in the vertical temperature gradient? Does it make physical sense based on data and modeling? Does it make physical sense in terms of heating and cooling? Constraints on upper limits in accretion heating??}

\begin{figure*}
    \centering
    \includegraphics[width=0.8\textwidth]{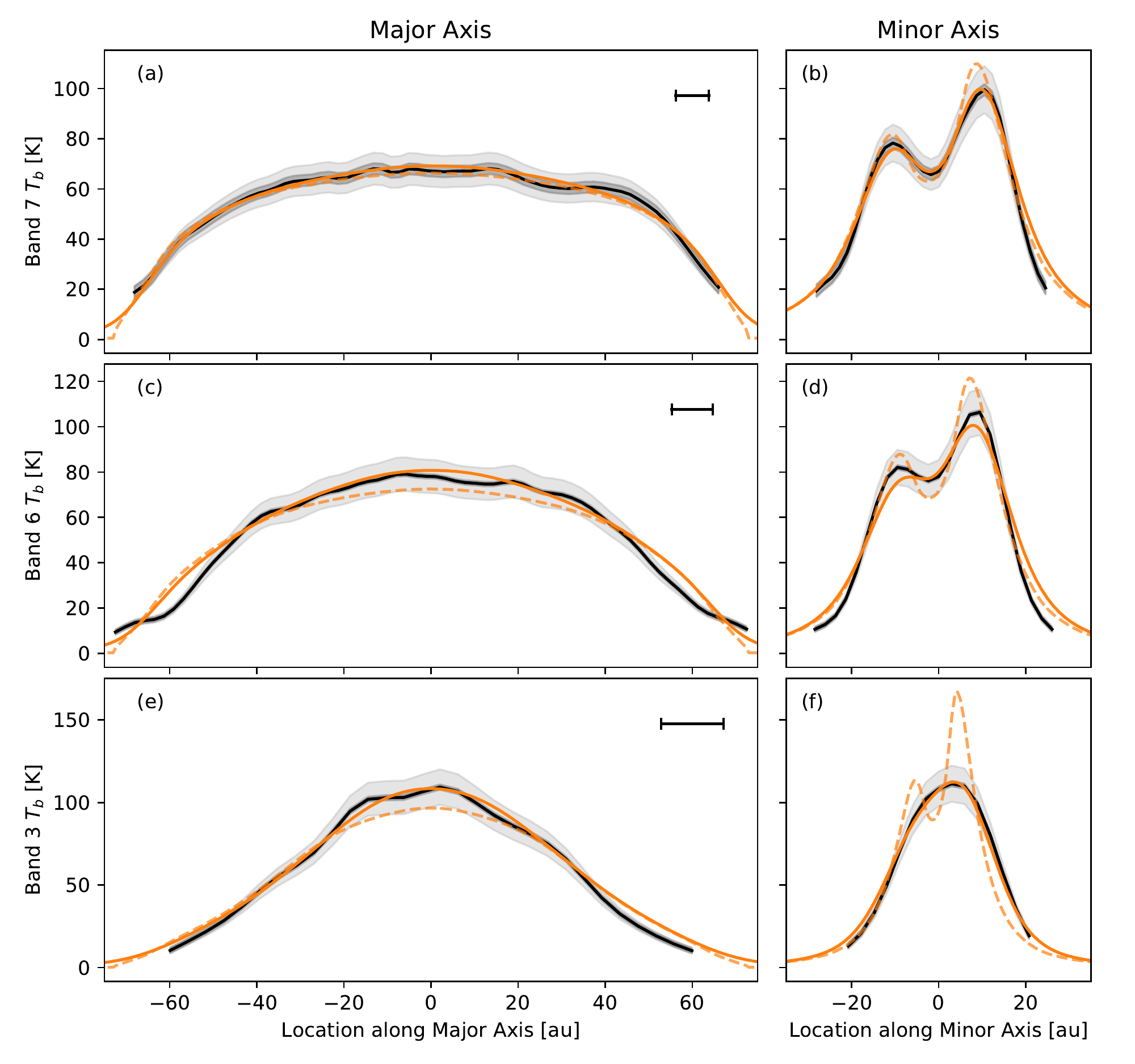}
    \caption{
        The brightness temperature along the major (left column) and minor (right column) axes of the disk. The solid black lines are the observations with the shaded regions representing the uncertainties associated with noise (darker shade) and absolute flux calibration (lighter shade), respectively. The dashed red lines are the model results before convolution, while the solid red lines are results after convolution. The cuts at different wavelengths go from top (Band 7) to bottom (Band 3). 
            }
    \label{fig:model_fit}
\end{figure*}

We find the best fit parameters by fitting the major and minor axis cuts with the same method described in Section \ref{ssec:application}. The best fit parameters are shown in Table \ref{tab:axisym_par}. We plot the brightness temperature of the convolved major and minor axes cuts in Fig. \ref{fig:model_fit} to compare to the observations. Across all three wavelengths, the convolved models can broadly reproduce the observations. To understand the effects of finite resolution, we also show the non-convolved model. We will first examine the minor axis cuts, then discuss how the major axis cuts are affected by considering the inclined 2D model.

\begin{figure*}
    \centering
    \includegraphics[width=\textwidth]{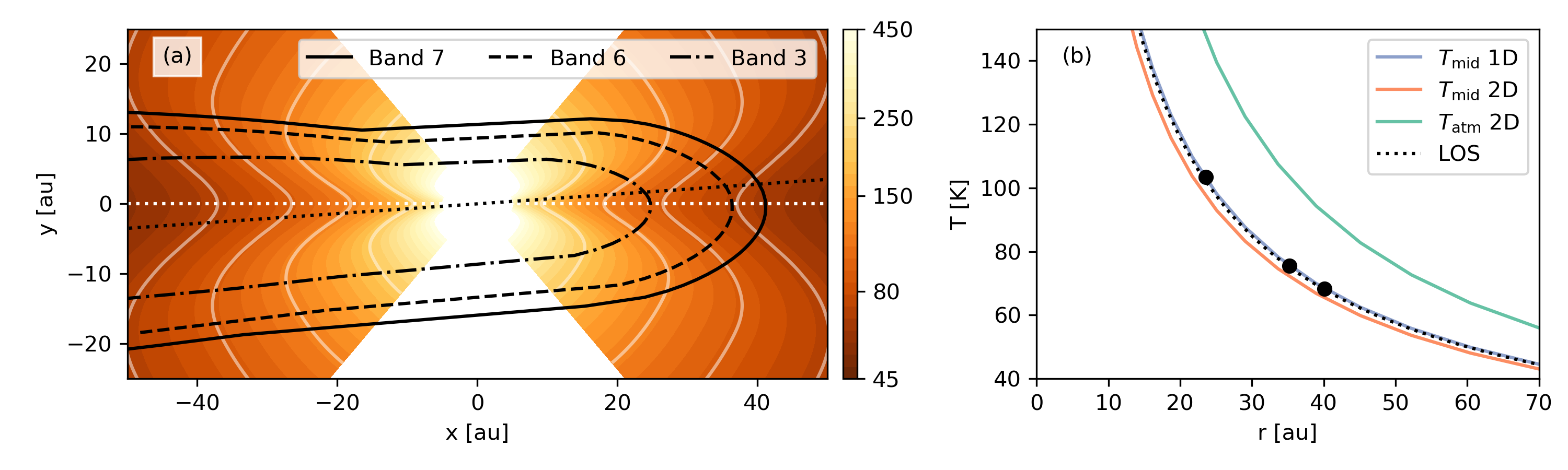}
    \caption{
        Panel a: the best-fit temperature distribution in the meridian plane (color map), with white contours of 70, 90, 120, and 200 K respectively. Positive $x$ is closer to the observer and the white dotted line marks the midplane. The $\tau'=1$ surfaces for Bands 7, 6, and 3 are plotted in black solid, dashed, and dash-dotted lines. The black dotted straight line represents the line of sight that crosses through the position of the central source. Panel b: the fitted temperature profiles as a function of spherical radius. $T_{\text{mid}}$ 1D is the temperature from the semi-analytical model. The midplane ($T_{\text{mid}}$ 2D) and atmospheric ($T_{\text{atm}}$ 2D) profiles for the 2D model using Eq.~(\ref{eq:temp2d_Tmid}) and Eq.~(\ref{eq:temp2d_Tatm}). The temperature along the black dotted line of sight in Panel a is plotted as a black dotted line (LOS). The black filled circles are the temperature at the $\tau'=1$ surfaces of the three wavelengths that crosses the line of sight. 
            }
    \label{fig:model_struc}
\end{figure*}

In Fig. \ref{fig:tbcut}b, we showed the brightness temperature along the minor axis of the disk at multiple wavelengths. The key features that need to be explained are the asymmetric double peak, the width of the profile, and the wavelength dependence of the profiles (i.e., the width decreases and the peak increases with increasing wavelength). To understand the minor axis cuts, we plot the inferred meridian temperature structure in Fig. \ref{fig:model_struc}a with the $\tau'=1$ surfaces of the three wavelengths as seen by an observer. The $\tau'=1$ surfaces are only plotted for line of sights with total optical depths larger than 1. Similar to the midplane case (Fig. \ref{fig:semiana_opt_normmax}a), the $\tau'=1$ surface for shorter wavelengths trace the outer regions, since the opacity is higher, or equivalently, $\tau_{0,\nu}$ is larger. With decreasing $\tau_{0,\nu}$, the width of the minor axis profile decreases, since the region of the disk seen by the observer that is optically thick becomes smaller, which is similar to the major axis case in Section \ref{sec:semianalytical}. 

The characteristic optical depth $\tau_{0,\nu}$ determines the width of the minor axis profile, or specifically, the extent of the $\tau'=1$ surface seen by the observer given a vertical density distribution. The vertical density distribution itself though is determined by the stellar mass and the vertical temperature structure through the vertical hydrostatic equilibrium. With a larger stellar mass or lower temperature, the material is more confined to the midplane and thus decreases the width of the profile for a given $\tau_{0,\nu}$. 

The double-peaked profile of the minor axis cuts is caused by vertical temperature gradient seen by the observer in the near-side of the disk ($x > 0$). In Fig. \ref{fig:model_struc}a, the $\tau'=1$ surface traces the colder midplane near the center of the profile and the warmer atmosphere near the outer wings. The central dip thus corresponds to the midplane temperature and the peaks correspond to the atmospheric temperature of the near side of the disk. Beyond the two peaks of the minor axis profile, the $\tau'=1$ surface happens in the far side of the disk ($x < 0$), which means the near side is optically thin for those line of sights, and thus the outer wings depend on the tenuous atmosphere. The distance between the two peaks is determined by $h_{s}$ and $\tau_{0,\nu}$. With larger $h_{s}$, the vertical location where the warm $T_{\text{atm}}$ occurs increases ($h_{s}R$; see Eq. (\ref{eq:temp2d})) and thus increases the distance between the two peaks. On the other hand, when $\tau_{0,\nu}$ increases, the $\tau'=1$ surface moves outward to where $h_{s}R$ is larger. 

The asymmetry of the double peak is due to the inclination, since the $\tau'=1$ surface above the midplane ($y > 0$ in Fig. \ref{fig:model_struc}a) reaches a smaller radii where the temperature is higher, whereas the $\tau'=1$ surface below the midplane ($y < 0$) reaches a larger radii. Since the brightness temperature profile at one wavelength is determined by the temperature traced by a $\tau'=1$ surface and the location of the $\tau'=1$ surface is determined by $\tau_{0,\nu}$, the brightness temperatures at multiple wavelengths measures the atmospheric temperature at multiple locations, analogous to the major axis cuts. Thus, with $R_{0}$ mainly determined by the major axis cuts, one can determine $T_{0,\text{atm}}$ from the values of the brightness temperature and $q_{\text{atm}}$ by using multiple wavelength.

For the major axis cuts, the model produces the same behavior we have discussed from the semi-analytical model, e.g., a decrease in profile width with smaller $\tau_{0, \nu}$, an increase in peak brightness temperature, and the change in ``boxiness." This is expected since the inclination is nearly $90^{\circ}$. There are slight differences between the results from the 1D semi-analytical model and that from the major axis cuts of the 2D model that affect the best fit parameters. 

First, finite spatial resolution averages the dark lane and brighter atmospheric regions. The central peak of the convolved model is larger than that of the non-convolved model, because the emission from the hotter atmosphere is mixed into the midplane. Second, by including inclination, the plateau region of the profile for the ALMA Bands becomes more peaky. This is because the region that is traced by the $\tau'=1$ surface is slightly above the midplane where the temperature is higher (see the line of sight, black dotted line, in Fig. \ref{fig:model_struc}a).

Since beam averaging and inclination makes the major axis cuts more peaky, the $\tau_{0,\nu}$ at Bands 7, 6, and 3 found from the 2D model are higher than those derived from the semi-analytical model in order to produce flatter intrinsic profiles that match the observed ones after the beam averaging and inclination effects. Thus, the $\tau_{0,\nu}$ from the semi-analytical model serves as a lower limit. Additionally, the resolved double peak feature along the minor axis for Bands 7 and 6 also requires higher $\tau_{0,\nu}$ than what was obtained in the 1D case. The $\tau_{0,\nu}$ found from the 2D model is thus constrained by both the major and minor axes which the semi-analytical model did not consider. Since the temperature structure is constrained by the brightness temperature profiles both in the major and minor axes with an uncertainty of $\sim 10\%$ for the absolute flux uncertainty and the stellar mass is measured from the gas kinematics with an uncertainty of $\sim 20\%$ \citep{Lee2017_com}, the vertical density distribution and thus $\tau_{0,\nu}$ is likely constrained to within $30\%$ as a conservative estimate of the systematic uncertainty. 

\begin{table}
    \centering
    \begin{tabular}{l c c}
        Parameter & Variable & Value \\
        \hline
        Stellar mass [$M_{\odot}$] & $M_{*}$ & 0.25 \\
        Inclination [$^{\circ}$] & $i$ & 86\\
        Disk Edge [au] & $R_{0}$ & 66.0 $\pm$ 0.8 \\
        Midplane temperature at $R_{0}$ [K] & $T_{0, \text{mid}}$ & 45 $\pm$ 2 \\
        Midplane temperature power-law index & $q_{\text{mid}}$ & 0.75 $\pm$ 0.09 \\
        Atmospheric temperature at $R_{0}$ [K] &  $T_{0, \text{atm}}$ & 59 $\pm$ 5 \\
        Atmospheric temperature power-law index & $q_{\text{atm}}$ & 0.89 $\pm$ 0.07 \\
        Dimensionless height where $T_{\text{atm}}$ begins & $h_{s}$ & 0.6 $\pm$ 0.3 \\
        sharpness of the transition & $\delta$ & 2 \\
        Characteristic Optical Depths & $\tau_{0, \text{B7} }$ & 1.2 $\pm$ 0.1 \\
            & $\tau_{0,\text{B6}}$ & 0.85 $\pm$ 0.07 \\
            & $\tau_{0,\text{B3}}$ & 0.32 $\pm$ 0.04

    \end{tabular}
    \caption{
        The best fit values and the fixed parameters (those without uncertainty) for the 2D model. 
            }
    \label{tab:axisym_par}
\end{table}

\begin{figure}
    \centering
    \includegraphics[width=\columnwidth]{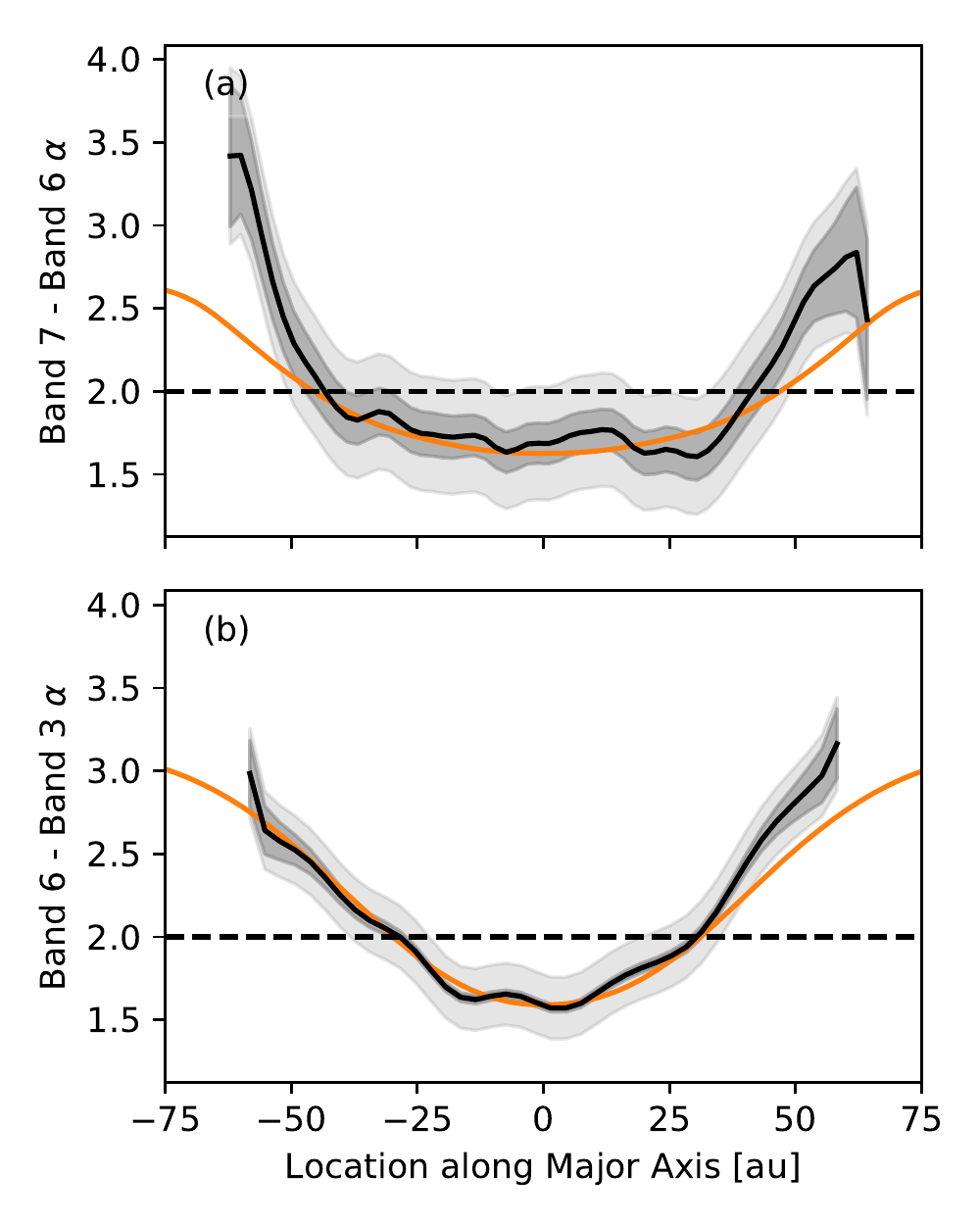}
    \caption{
        Spectral index between Bands 7 and 6 (panel a) and between Bands 6 and 3 (panel b). The black solid line is the measured value with the dark shaded region as the uncertainty due to noise and the light shaded region as the uncertainty from a $10\%$ absolute flux uncertainty. The orange line is the spectral index from the best fit model. 
            }
    \label{fig:model_spec}
\end{figure}

From the best fit parameters, we calculate the spectral index between consecutive wavelengths using the beam size of the lower angular image as was done for the observed spectral index images. Fig. \ref{fig:model_spec} show the spectral index from the best fit model compared to the measured spectral index along the major axis. The broad agreement between the model and data lends support to the notion that low spectral index ($\alpha < 2$) can indeed be produced by an increasing temperature away from the observer along the line of sight which, in this nearly edge-on case, is due to the radial temperature gradient in the disk. We should caution the reader that scattering can lower the spectra index below $2$ as well, although it is not required in this case.

\begin{figure}
    \centering
    \includegraphics[width=\columnwidth]{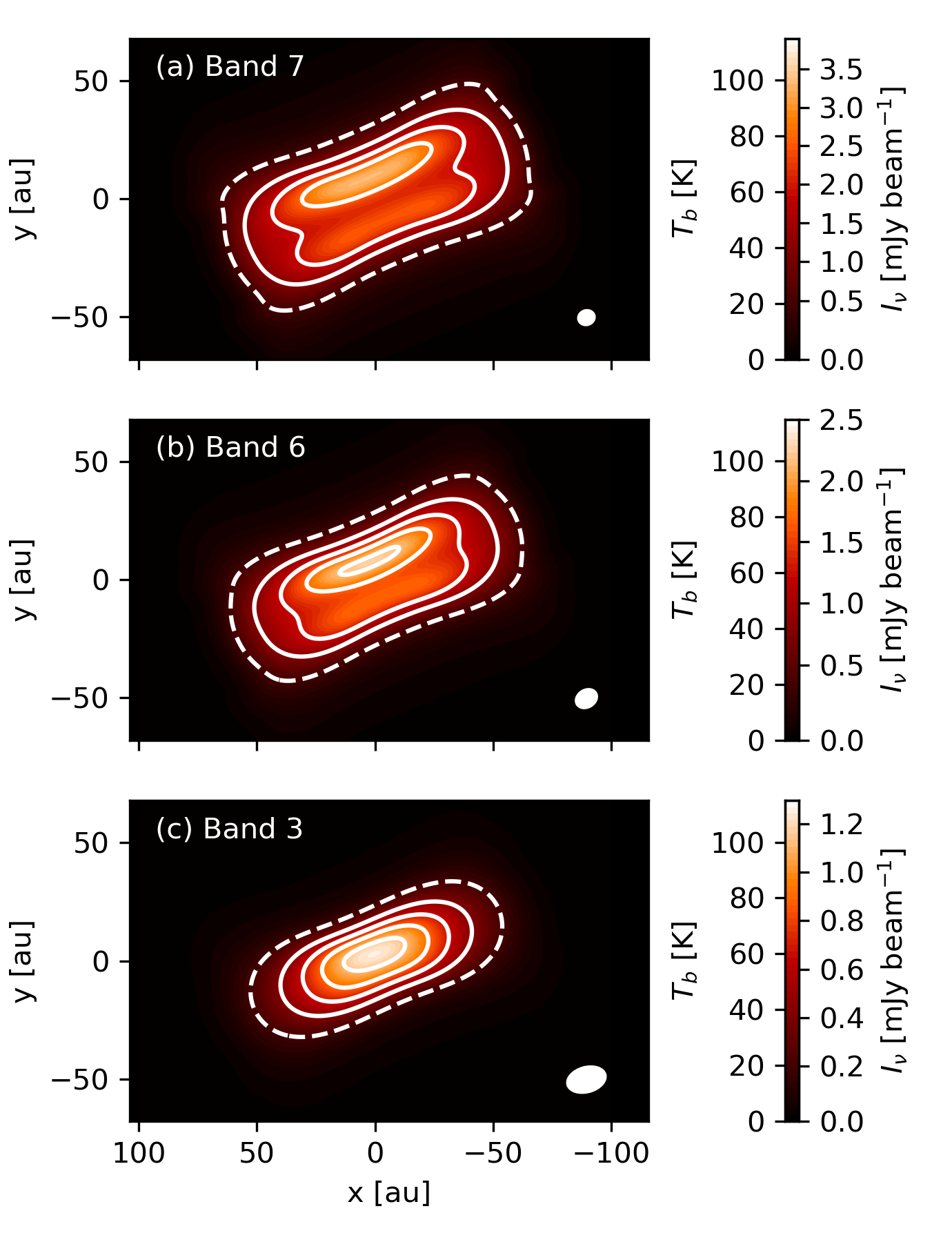}
    \caption{
        The multi-wavelength images from the model convolved by a beam size of the corresponding observations shown as white filled ellipses in the lower right of each panel. The colormap is plotted in brightness temperature. The contours are brightness temperature levels starting from 20K (the dashed contour) in steps of 20. 
            }
    \label{fig:model_image}
\end{figure}

In Fig. \ref{fig:model_image}, we plot the convolved model images. Though the major and minor axes fit the data quantitatively, there appears to be discrepancies in the general two-dimensional morphology especially for Bands 7 and 6 images. At the outer boundaries of the disk (e.g. the 20K contour), the extent of the model images is more rectangular than the observations (which resembles more like a ``football"). \cite{Lee2017_lane} proposed that the scale height of the disk decreases with radius beyond $R \sim 36 \pm 8$au towards the edge though the physical explanation is still unknown. One possibility is that the infalling material from the envelope collides with the surface of the outer edge and thus provides an additional ram pressure to decrease the dust scale height (see e.g., Fig. 19 of \citealt{Li2014}). %Furthermore, as $Q \rightarrow 1$, radial and vertical self-gravity becomes increasingly important, thus the current 1+1D model that only includes gravity from the central point source is not by itself consistent with $Q=1$. At least a 2D model (instead of our 1+1D model prescribed in Section \ref{sec:axisymmetric}) that includes both the disk and the envelope and treats self-gravity (both vertically and radially) will be required to verify the possibility. 
More refined disk models including the potential dynamical effects of the envelope accretion are needed to resolve this discrepancy.
%detailed modeling of high angular resolution images of the dust and of the gas can shed light onto the complexities of these young and dynamic sources such as HH 212 mms. 

We should stress that the dust opacities inferred from the model fitting above are absorption opacities obtained under the assumption of a negligible contribution from scattering. We briefly explore the effects of scattering in the Appendix \ref{sec:scattering}, where we demonstrate that scattering tends to increase the characteristic temperature $T_0$ and the total (extinction) opacity inferred based on the peak value and shape of the observed brightness temperature distribution.

% ===========================================================
\section{Discussion} \label{sec:discussion}

% opacity constraint
% - optically thick even at Band 3
% - grain size constraint
\subsection{Support for the Beckwith et al. opacity prescription} 
\label{ssec:opacity}

%\red{Need a standalone figure for the inferred opacities as a function of wavelength, both per unit gas and per unit dust assuming dust-to-gas ratio of 0.01. This is the main product of the paper independent of the comparison to dust opacity models! Recap how the opacities are determined and whether it is robust or not.}

\begin{figure*}
    \centering
    \includegraphics[width=0.7\textwidth]{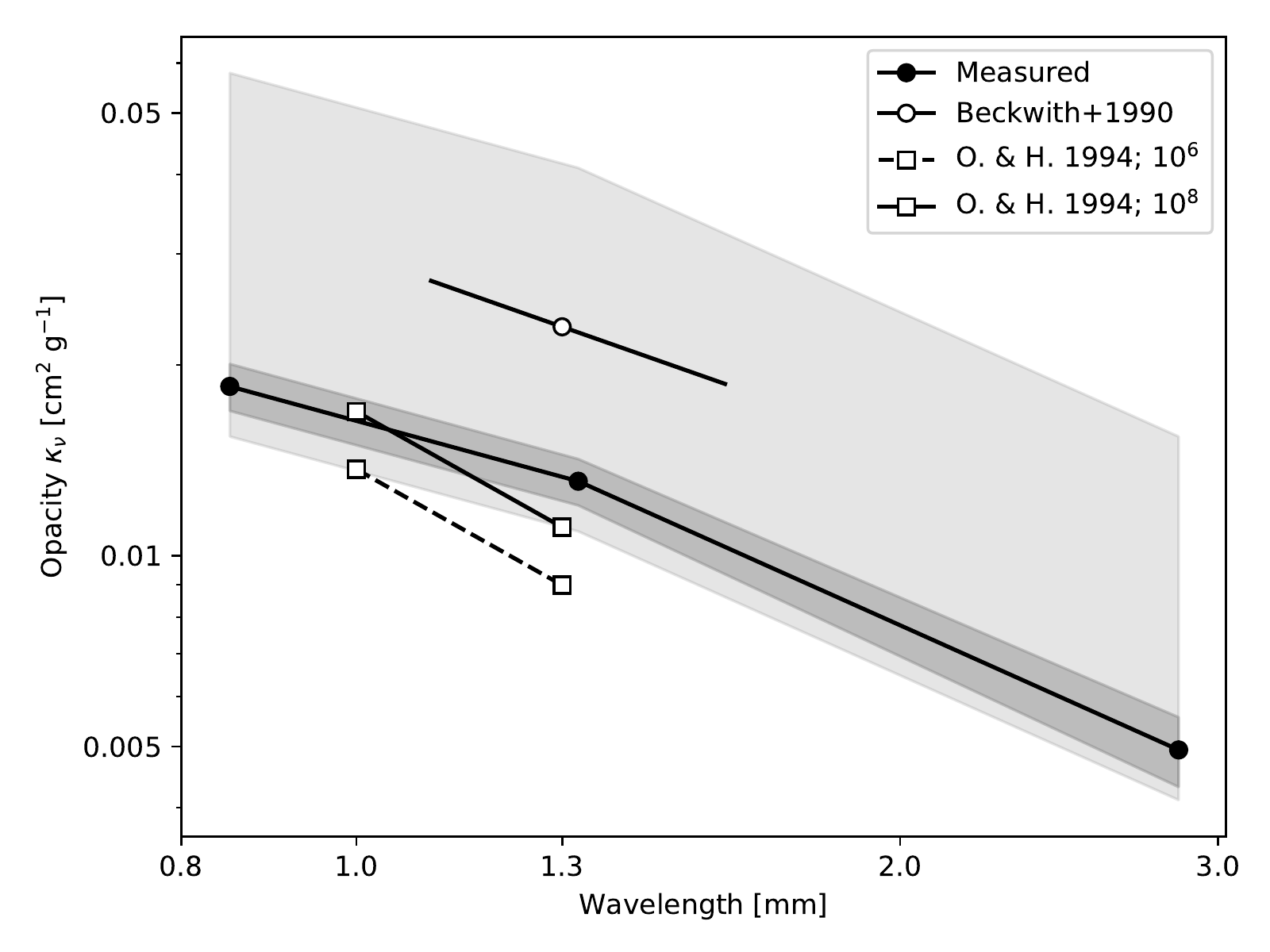}
    \caption{
        Dust opacities (absorption cross section per gram of gas and dust) inferred in the HH 212 protostellar disk as a function of wavelength at ALMA Band 7, 6 and 3. The lighter shaded region denotes the estimated uncertainties associated with the Toomre $Q$ parameter and the stellar mass $M_{*}$. The darker shaded region shows the estimated uncertainties due to noise and assumes $Q=1$ and $M_{*}=0.25 M_{\odot}$. The open circle is the \protect\cite{Beckwith1990} opacity at 1.3 mm with a black solid line segment to show its adopted $\beta=1$. The open squares are opacities from \protect\cite{Ossenkopf1994} at 1 and 1.3 mm assuming a gas-to-dust mass ratio of 100. The open squares with a solid black line is for dust with an MRN distribution with thin ice mantles for a gas number density of $10^{8}$ cm$^{-3}$, while the open squares with a dashed black line is for a gas number density of $10^{6}$ cm$^{-3}$.
            }
    \label{fig:opacities}
\end{figure*}

One of our most important results is the constraint on the dust opacity $\kappa_\nu$. As discussed in the introduction, the dust opacity is important because it is the key to convert the observed flux of dust emission to the mass. In our proposed new method, it is determined from the characteristic optical depth $\tau_{0,\nu}$ (see Eq.~(\ref{eq:op2})), which is in turn determined mostly from the shape of the observed brightness temperature profile along the major axis (see Fig.~\ref{fig:semiana_opt_normmax}c). The opacity (per gram of gas and dust, not just dust) is plotted in Fig.~\ref{fig:opacities} as a function of wavelength for ALMA Band 7, 6, and 3 for the fiducial values of the Toomre parameter $Q=1$ and stellar mass $M_*=0.25$~M$_\odot$. The noise uncertainty is shown as a dark shaded region. The absolute uncertainties are dominated by $Q$ and $M_*$, which enter the opacity determination through $\kappa_\nu\propto Q/M_*$. In particular, if the HH 212 disk is marginally gravitationally unstable, $Q$ can be as large as $2.5$ \citep{Kratter2010}, which would increase the opacity by a factor of $2.5$ over the fiducial value. Since the stellar mass is determined to $M_*=0.25\pm0.05$~M$_\odot$ \citep{Lee2017_com}, the opacity can additionally vary by $20\%$. The range of $\kappa_{\nu}$ due the uncertain range of $Q$ and $M_*$ is shown in the lighter shaded regions in Fig.~\ref{fig:opacities}.  

%From the opacities inferred for the 3 wavebands, it is easy to compute the opacity indices ($\beta$) between adjacent bands, which are shown in Fig.~\ref{fig:opacities}b. The uncertainties on $\beta$ are smaller than those on the opacity $\kappa_\nu$, because it depends only on the ratio of the characteristic optical depths between two adjacent wavelengths, which are well constrained from the shape of the brightness temperature profile. Based on the semi-analytical model, we estimate that the uncertainty on $\kappa_\nu$ at a given wavelength is about $5\%$ (see Table~\ref{tab:curvefit}), which yields an uncertainty on $\beta$ that is order $10\%$ at most. This estimate is shown as a shaded region in Fig.~\ref{fig:opacities}b. Also plotted in the figure is the best fit value of $\beta$ across all 4 wavebands. 

For comparison, we plot in Fig.~\ref{fig:opacities} the widely used dust opacity prescription advocated by \cite{Beckwith1990} for young star disks: $\kappa_{\nu} = 0.1 (\nu / 10^{12} \rm{Hz})^{\beta}$ cm$^2$ g$^{-1}$ with $\beta=1$, or equivalently $\kappa_\lambda=2.3\times 10^{-2} (1.3 \rm{mm}/\lambda)$ cm$^2$ g$^{-1}$. The reference opacity of $2.3\times 10^{-2}$~cm$^2$~g$^{-1}$ at 1.3~mm (shown as open circle in Fig.~\ref{fig:opacities}) is remarkably close to, but about a factor of 1.8 times larger than, the minimum opacity of \kapSix~cm$^2$~g$^{-1}$ (corresponding to $Q=1$) that we inferred at the same wavelength (ALMA Band 6). Just as remarkably, the opacity index that we inferred for the (millimeter/submillimeter) ALMA Bands, $\beta = 1.06 \pm 0.15$ (noise uncertainty), is very close to that advocated by \cite{Beckwith1990}. We view these agreements as an independent confirmation of the \cite{Beckwith1990}'s prescription, which corresponds to a Toomre parameter of $Q\approx 1.8$ for the HH 212 disk, a value that is quite reasonable for a marginally gravitational unstable disk \citep{Kratter2010, Kratter2016}. 

Another oft-used dust opacity is that computed by \cite{Ossenkopf1994}. For grains coated with thin ice-mantles that evolved for $10^5$ years at a gas number density of $10^6$~cm$^3$ that is most commonly used in the literature \citep[e.g.][]{Tychoniec2018}, they obtained an opacity of $8.99\times 10^{-3}$~cm$^2$ g$^{-1}$ at 1.3~mm (assuming a gas-to-dust mass ratio of 100; open square with a dashed line in Fig.~\ref{fig:opacities}), which is a factor of $1.5$ below our inferred minimum value at the same wavelength. We do not favor this value of opacity, since it would formally imply a Toomre parameter of $Q\approx 0.7$ for the HH 212 disk, which is less physically justifiable \citep{Kratter2016}. The opacity of the same dust model, but calculated for a gas number density of $10^{8}$~cm$^{3}$ lies just within our range of uncertainty due to $Q$ and $M_{*}$. 
%Thus, computing opacities applicable for the denser young circumstellar disk environments may be beneficial. 
However, the same Ossenkopf \& Henning model predicts a spectral index of $\beta\approx 1.7$ between 0.7 and 1.3~mm, which is steeper than what we inferred between the three ALMA bands ($\beta\approx 1.1$). The disagreement is perhaps not too surprising because the \cite{Ossenkopf1994} opacity is formally computed for dense cores of molecular clouds where the grains may be different (e.g., smaller) than those in the disks studied by \cite{Beckwith1990} and in this paper.

%
% what the values and related comments? 
%

\subsection{Implications of the Inferred Dust Opacities}
\label{ssec:OpacityImplications}

The physical origin of the (sub)millimeter opacity prescription advocated by \cite{Beckwith1990} and supported by our work remains uncertain. It is well known that it cannot be reproduced by small (sub-micron) grains of diffuse interstellar medium \citep[e.g.][]{Weingartner2001}, which under-predicts the value $\kappa_{1.3{\rm mm}}$ at the reference wavelength of 1.3~mm by an order of magnitude and over-predicts the opacity index (with $\beta\sim 2$ rather than $\sim 1$; see, e.g., Fig.~3 of \cite{Andrews2009} for an illustration). The relatively low value of $\beta\sim 1$ at (sub)millimeter wavelengths is often interpreted as evidence for large, mm or even cm-sized grains (\citealt{Draine2006}; see \citealt{Testi2014} for a review). Grain growth would also increase $\kappa_{1.3 {\rm mm}}$ over the diffuse ISM value, making it more compatible with that advocated by \cite{Beckwith1990}. 
However, we stress that both the opacity $\kappa_\nu$ and the opacity index $\beta$ can be strongly affected by the shape of the grains and particularly their fluffiness \citep{Wright1987}, which are not well determined observationally. In addition, for the HH 212 disk, it is unclear whether mm/cm sized grains would be compatible with the lack of evidence for dust settling (see Section \ref{ssec:NoSettling}) and with the detection of dust polarization at $\sim 1\%$ level in ALMA Band 7 \citep{Lee2018_pol}, since mm/cm grains are not effective at producing polarization in ALMA Band 7 through scattering \citep{Kataoka2015,Yang2016_HLTau} and such grains are too large to be aligned with the disk magnetic field because of a long Larmor precession timescale unless their magnetic susceptibility is enhanced by several orders of magnitude through, e.g., superparamagnetic inclusions (e.g., H. Yang 2020, submitted). In view of these uncertainties, it is prudent to refrain from drawing firm conclusions on grain size.

\subsection{Implications of a Warm Phase Early in Disk Formation and Evolution}

Besides dust opacity, another interesting finding of our paper is that the HH 212 disk is rather warm. It has a midplane temperature of $\sim 45$~K near the disk outer edge at a radius of $\sim 70$~au. The temperature increases to $\sim 70$, $\sim 75$ and $\sim 100$~K at the $\tau'=1$ surfaces for ALMA Band 7, 6 and 3, which are located at a radius of $40$, $35$ and $25$~au, respectively. The temperature is even higher near the disk surface, in order to produce the striking ``hamburger-shaped" morphology with a dark lane sandwiched by two brighter layers that is observed in Band 7 and 6. It is therefore reasonable to conclude that the temperature is well above the sublimation temperature of CO ice ($\sim 20$~K) everywhere in this early disk\footnote{The higher temperature in the HH 212 disk compared to typical Class II disks is not too surprising because it has a rather high luminosity ($\sim 10\ L_\odot$), a massive envelope which can heat up the disk through back warming, and a much higher mass accretion rate ($\sim 5\times 10^{-6}$~M$_\odot$~yr$^{-1}$, \citealt{Lee2014}) which can provide a much stronger accretion heating. There are, however, uncertainties in the structure of the system, particularly its envelope and outflow cavity, that make a first principle calculation of the disk temperature difficult. It is beyond the scope of this paper.}, which should have strong implications for the disk chemistry. The situation is similar to, but more extreme than, another deeply embedded, nearly edge-on, protostellar disk L1527, where \cite{vantHoff2018} inferred a disk midplane temperature of $\sim 30$~K on the 50~au scale from molecular lines. As stressed by these authors (and others, e.g., \citealt{Harsono2015}), one implication of the rather high temperature would be that the chemical composition of the disk may not be entirely inherited from the molecular cloud core, because the ices of low sublimation temperatures (such as CO and N$_2$) inherited from the collapsing core are expected to sublimate, causing an at least partial reset of the chemistry \citep[e.g.][]{Pontoppidan2014}. 

A warm phase early in the formation and evolution of disks indicated by our observations and analysis of HH 212 mms also has implications for the timing of the onset of CO depletion from the gas phase that is widely inferred in older protoplanetary disks \citep{Bergin2018}. We do not expect such large CO depletion in young warm HH 212 mms-like disks, for two reasons. First, without a cold midplane where CO can condense onto the grain surface, it would difficult to remove the CO from the warmer gas layers by vertical mixing \citep{Kama2016, Krijt2018}. Second, the inferred disk temperature of $\sim 45$~K (or higher) is also well above the optimal temperature of $\sim 20-30$~K found by \cite{Bosman2018} for efficient chemical depletion of the CO; in particular, chemical depletion of CO by hydrogenation on the grain surface to form species such as CH$_3$OH would not work efficiently\footnote{An implication of the warm HH 212 disk is that its observed CH$_3$OH and especially CH$_2$DOH \citep{Lee2017_com} are likely carried into the disk by the collapsing protostellar envelope rather than being produced on the grain surface locally. This is in line with the conclusion of \cite{Drozdovskaya2020} that deuterated methanol in comet 67p/Churyumov-Gerasimenko has an origin in the cold pre-stellar phase of solar system formation.}. Whether this expectation is met or not is important but remains to be determined \citep{Lee2017_com, Lee2019}. If so, it would support a transition to a CO-depleted disk somewhere between late Class 0 phase (as represented by HH 212 mms and possibly L1527) and early Class II phase, as indicated by the recent observations of \cite{Zhang2020} and, furthermore, that the transition may happen as a result of the reformation of the CO freeze-out region near the midplane as the disk cools down with time.

\subsection{No Vertical Dust Settling Required in the HH 212 Disk}
\label{ssec:NoSettling}

A dust grain elevated above the dust midplane experiences the gravitational pull of the central star in the vertical direction and is thus expected to settle towards the midplane of the disk over time. This expectation is confirmed observationally in a number of evolved, Class II disks (see also \citealt{Villenave2020}), such as IM Lup and 2MASS J16281370-2431391 (``the Flying Saucer"). For IM Lup, the image observed at millimeter with ALMA \citep[e.g.][]{Andrews2018} shows a fairly flat disk with clear spirals, while the image observed in scattered light with the Spectro-Polarimetric High-contrast Exoplanet REsearch (SPHERE) on the Very Large Telescope array (VLT) in optical shows a disk surface composed of smaller grains that is well above the midplane \citep{Avenhaus2018}. For the ``Flying Saucer," modeling the millimeter-sized dust grains in \cite{Guilloteau2016} shows a dust scale height that is roughly half the scale height of micron-sized dust grains seen in infrared \citep{Grosso2003} and also of the gas \citep{Dutrey2017}. This settling of large grains into a thin sub-layer within the disk is thought to play a crucial role in planet formation \citep[e.g.][]{Dubrulle1995, Dullemond2004, Testi2014}. The timing of the onset of significant dust settling in the process of disk formation and evolution remains, however, uncertain. 

% various disks have evidence of dust settling: Flying Saucer (dust vs gas), IM Lup (Sphere vs alma)
The dust in the young, deeply embedded HH 212 disk does not appear to be settled. From fitting the data with an axisymmetric disk in hydro-static equilibrium, we find that dust settling is not required to approximate the resolved images in Band 7, 6, and 3 along the minor axis. In particular, the dust disk still needs to be optically thick in the atmosphere to trace the vertical temperature gradient and create the double peak which is difficult if dust is settled like Class II disks. The lack of dust settling is not too surprising since the HH 212 disk is still receiving from the massive infalling envelope dust grains (and gas) that are presumably relatively small and the age of the system, estimated to be roughly $3\times 10^4$~years \citep{Lee2014}, may be too short for the grains to grow large enough for significant dust settling to occur. In addition, the inferred high mass accretion rate (of order $\sim 5\times 10^{-6}$~M$_\odot$~yr$^{-1}$, \citealt{Lee2014}) may require a high level of turbulence, which may make the dust settling more difficult, although the mass accretion can also be driven by a disk-wind, for which there is some tantalizing evidence \citep{Lee2018_shell}. The level of turbulence can potentially be constrained by molecular line observations, particularly using ALMA in the upcoming Band 1 which is expected to be mostly optically thin. 
%
%If the level of turbulence can be obtained from future observations, it can provide an important piece to whether or not dust settling can occur and also the physical mechanism driving the turbulence (e.g., magnetorotational instability and vertical shear instabilities), since various mechanisms predict a different vertical profile of turbulence \citep{Balbus1991, Balbus1998, Nelson2013, Flaherty2017}. Since the continuum disk at Band 3 is still optically thick, line observations using ALMA at Band 1 will be particularly suitable. 
%
In any case, if the dust has yet to settle in the HH 212 disk, it would point to significant dust settling happening some time between the late Class 0 phase (represented by HH 212 mms) and Class II phase, similar to the onset of CO depletion discussed above. The transition may happen due to a combination of a longer time for grains to grow and settle and disks becoming less turbulent over time, which is consistent with the trend of decreasing mass accretion rate with time \citep[e.g.][]{Yen2017}. 

\section{Conclusions} \label{sec:conclusions}

We presented high-resolution multi-wavelength ALMA and VLA dust continuum observations of the nearly edge-on Class 0 disk HH 212 mms and modeled the data quantitatively. Our main results are as follows:

\begin{enumerate}[label=\arabic*)]
    \item New ALMA dust continuum images at Bands 9, 6, and 3 are presented with angular resolutions of $\sim$70, 23, and 36 mas, respectively. The shortest wavelength Band 9 emission traces both the disk and the envelope. The dark lane along the disk midplane first resolved in previous Band 7 observations is well resolved at Band 6 as well, but less so at Band 3. The disk is less well-resolved at the longest wavelength of the VLA Band Ka, which has a resolution of $\sim$58 mas. For the best resolved ALMA Bands 7, 6 and 3, we find a clear trend that, as the wavelength increases, the brightness temperature profile along the major axis becomes narrower, with a higher peak value towards the center (see Fig.~\ref{fig:tbcut}). In addition, the spectral indices between adjacent wavebands have values below $3$ over most of the disk, with unusually low $\alpha < 2$ in the central part between ALMA bands. 
    
    %{(from temperature gradient?)} 
    
    %
    % qualitative principle, shape to tau_0, flux to temperature,  
    %
    % \alpha < 2 from temperature gradient
    %
    \item We developed a simple 1D semi-analytical model to illustrate how to use the observed brightness temperature ($T_b$) profile along the major axis of an edge-on disk to determine its midplane temperature as a function of radius and the dust opacity $\kappa_\nu$ under the assumption of negligible scattering. From the degree of peakiness of the $T_b$ profile one can infer the dimensionless characteristic optical depth $\tau_{0,\nu}=\rho_0 R_0\kappa_\nu$ (where $\rho_0$ and $R_0$ are the density and radius at the disk outer edge), which controls the location of the sight line where the fainter optically thin outer part of the disk transitions to the brighter inner part (see Fig~\ref{fig:semiana_opt_normmax}). From the peak brightness along the optically thick sight line towards the center, one can infer the temperature at the $\tau'=1$ surface, whose distance from the origin (i.e., radius) is mostly controlled by $\tau_{0,\nu}$. Gravitational stability considerations put an upper limit on $\rho_0$ which, combined with the observationally inferred characteristic optical depth $\tau_{0,\nu}$ and disk outer radius $R_0$, yields a lower limit on the dust opacity $\kappa_\nu$. We find that the unusually low spectral index of $\alpha < 2$ observed in the central part of the HH 212 disk can naturally be explained by the inferred temperature increase towards the center, because the longer wavelength probes a warmer region closer to the origin. 
  
  %
  % 2D model results
  %
    \item Building on the 1D semi-analytical model, we numerically fitted the multi-wavelength dust continuum observations of the HH 212 disk in two dimensions, accounting for a small deviation from exact edge-on orientation and telescope beam convolution that blends the emission from the disk surface and midplane. We inferred a temperature of $T_0\approx 45$~K at the disk outer edge $R_0\approx 70$~au, which increases radially inward roughly as $T(R)\propto R^{-0.75}$ along the midplane and vertically away from the midplane. We inferred a characteristic optical depth of $\tau_{0,\nu}$ of $\sim 1.2$, 0.85, and 0.30 for ALMA Band 7, 6, and 3 (see Table~\ref{tab:axisym_par}), corresponding to a dust opacity (per gram of gas and dust) of $\kappa_{\nu} \sim $ \kapSeven, \kapSix, and \kapThree $\times Q (0.25 M_\odot / M_*)$~cm$^2$~g$^{-1}$, respectively, where $Q$ is the Toomre parameter and $M_*$ the stellar mass.  
    
    %
    % Implications, kappa, T, no settling
    %
    \item Our inferred dust opacities lend support to the widely used prescription $\kappa_\lambda=2.3\times 10^{-2} (1.3 {\rm mm}/\lambda)$ cm$^2$~g$^{-1}$ advocated by \cite{Beckwith1990}. It corresponds to a Toomre parameter of $Q\approx 1.8$ for the HH 212 disk, which is physically reasonable for a marginally gravitationally unstable disk. Our inferred disk temperature of $\sim 45$~K and more is well above the sublimation temperatures of ices such as CO and N$_2$, which supports the notion that the disk chemistry cannot be completely inherited from the protostellar envelope. The relatively high temperature also makes it difficult to deplete CO from the gas phase in the early, deeply embedded stage of disk formation and evolution either physically through freeze-out or chemically. In addition, no dust settling is required in fitting the multi-wavelength HH 212 dust continuum data, indicating that the grains accreted from the massive envelope have yet to grow large enough or the fast accreting disk is too turbulent for the dust to settle or both. 
\end{enumerate}

\section*{Acknowledgements}
We express our gratitude to the National Radio Astronomy Observatory (NRAO) F2F support team in Charlottesville, Virginia, especially T. Booth, for their helpful and dedicated support. We thank C.P. Dullemond for making \textsc{RADMC-3D} publicly available and for providing insightful help. We also thank D. Segura-Cox, H. Yang, and C.-Y. Lai for fruitful discussions and the referee for constructive comments. Z.-Y.D.L. acknowledges support from ALMA SOS (SOSPA7-001), Ministry of Science and Technology of Taiwan (MoST 104-2119-M-001-015-MY3), and NASA 80NSSC18K1095. C.-F.L. acknowledges grants from the Ministry of Science and Technology of Taiwan (MoST 104-2119-M-001-015-MY3 and 107-2119-M-001-040-MY3) and the Academia Sinica (Investigator Award AS-IA-108-M01). ZYL is supported in part by NASA 80NSSC20K0533 and NSF AST-1716259 and 1910106. Parts of this research were carried out at the Jet Propulsion Laboratory, California Institute of Technology, under contract 80NM0018D0004 with the National Aeronautics and Space  Administration. This paper makes use of the following ALMA data: ADS/JAO.ALMA\#2012.1.00122.S, ADS/JAO.ALMA\#2015.1.00024.S, ADS/JAO.ALMA\#2017.1.00044.S, ADS/JAO.ALMA\#2017.1.00712.S. ALMA is a partnership of ESO (representing its member states), NSF (USA) and NINS (Japan), together with NRC (Canada), MOST and ASIAA (Taiwan), and KASI (Republic of Korea), in cooperation with the Republic of Chile. The Joint ALMA Observatory is operated by ESO, AUI/NRAO and NAOJ. The National Radio Astronomy Observatory is a facility of the National Science Foundation operated under cooperative agreement by Associated Universities, Inc.

% We thank the DSHARP group and DIANA group for making the dust models publicly available and convenient to use. 

% Band 9: 2012.1.00122.S
% Band 7 Lee 2017: 2015.1.00024.S, 
% Band 7 pol: 2017.1.00044.S
% Band 6: 2017.1.00172.S
% Band 3: 2017.1.00172.S

%%%%%%%%%%%%%%%%%%%%%%%%%%%%%%%%%%%%%%%%%%%%%%%%%%

\section*{Data Availability}
The ALMA data can also be obtained from the ALMA Science Data Archive\footnote{\url{https://almascience.nrao.edu/asax/}}. The Band Ka data can be obtained from \url{https://dataverse.harvard.edu/dataverse/VANDAMOrion}. Additional data underlying this article is available from the corresponding author upon request. 



\bibliographystyle{mnras}
\bibliography{main} % if your bibtex file is called example.bib

%%%%%%%%%%%%%%%%%%%%%%%%%%%%%%%%%%%%%%%%%%%%%%%%%%

%%%%%%%%%%%%%%%%% APPENDICES %%%%%%%%%%%%%%%%%%%%%

\appendix

\section{Disk Center} \label{sec:diskcenter}

The central location of the disk, ideally the location of the star, is necessary to compare across different wavelengths which are observed at different times. However, for edge-on disks, the star is hidden making it difficult to locate the exact center. We utilize the Band 7 high angular resolution shown in this paper (on 2015 November when the longest baseline was used) and the Band 7 image observed in 2017 December. After convolving the higher resolution image to match resolution of the other image, we calculate the shift in position that minimizes the squared sum of the difference between the two images. We find a proper motion of $\sim 0.5, -2$ mas yr$^{-1}$ in RA and Dec. After correcting for proper motion, we adopt the peak of the image observed at Band Ka as the center. 
%However, based on comparisons to models in Section \ref{sec:axisymmetric}, we still needed a relative difference of 0.004" and 0.007" in RA and Dec between the ALMA and VLA images which could be due to absolute pointing error. 

\section{Derivations for peak brightness temperature and other properties} \label{sec:qobs}
In Section \ref{sec:semianalytical}, we derived the brightness temperature normalized by $T_{0}$ in the Rayleigh-Jeans limit as Eq.~(\ref{eq:tbsemi}). A simple relation exists for the peak brightness temperature and we demonstrate how the temperature gradient $q$ can be approximated by the peak brightness temperature across multiple wavelengths. 

With $\hat{b} \rightarrow 0$, the optical depth $\tau'$ (in normalized coordinates) becomes 
\begin{equation}
    \tau'(\hat{z}) = \dfrac{\tau_{0,\nu}}{ \hat{b}^{2} } \dfrac{1 + \dfrac{1}{2} \bigg(\dfrac{\hat{b}}{ \hat{z} }\bigg)^{2} - \dfrac{1}{2} \hat{b}^{2} - \text{sgn}(\hat{z}) }{ \sqrt{1 + \bigg( \dfrac{ \hat{b}}{ \hat{z}} \bigg)^{2}} }
\end{equation}
where $\text{sgn}$ is the sign function. For the optical depth to stay finite at $\hat{b}=0$, we are restricted to $\text{sgn}(\hat{z}) = 1$. Physically, it means the far half of the disk ($\hat{z} < 0$) is not seen at all which is expected since the midplane density becomes infinitely high as $\hat{z} \rightarrow 0$ and blocks all emission from the far half. We thus obtain 
\begin{equation} \label{eq:taub0}
    \tau'(\hat{z}) = \dfrac{\tau_{0,\nu}}{2} \bigg( \dfrac{1}{ \hat{z}^{2} } - 1 \bigg) \text{ , } \hat{z} > 0
\end{equation}
which is the same as simply integrating Eq. (\ref{eq:optdepth_p3}) assuming $\hat{z} > 0$: $\tau'(\hat{z}) = \tau_{0, \nu} \int_{\hat{z}}^{1} \hat{z}^{-3} d \hat{z}$. 
With $\tau'$ determined from Eq. (\ref{eq:taub0}), the peak brightness temperature is
\begin{align} \label{eq:Thatb0p3_derive}
    \hat{T}_{\nu}(\hat{b}=0) &=
        \int_{0}^{1} \dfrac{\tau_{0,\nu}}{ \hat{z}^{(q + 3)}}
        \exp{ \bigg[
            - \dfrac{\tau_{0,\nu}}{2} \bigg( \dfrac{1}{ \hat{z}^{2} } - 1 \bigg)
            } \bigg]
                d \hat{z} 
\end{align}
where $\sqrt{\hat{z}^{2}}$ is replaced by $\hat{z}$ for this integration range. We make a change in variable by defining 
\begin{align}
    u \equiv \bigg(\dfrac{\tau_{0,\nu}}{2} \bigg) \hat{z}^{-2}
\end{align}
and thus the relation to $d\hat{z}$ is 
\begin{align}
    du = - \tau_{0,\nu} \hat{z}^{-3} d \hat{z} 
\end{align}
integrating from $\infty$ to $\tau_{0,\nu} / 2$. 
By replacing all occurrences of $\hat{z}$ by $u$ in Eq. (\ref{eq:Thatb0p3_derive}) and organizing, we have 
\begin{align}
    \hat{T}_{\nu}(\hat{b}=0) &= 
        \bigg(\dfrac{\tau_{0,\nu}}{2} \bigg)^{- \dfrac{q}{2}} e^{ \dfrac{ \tau_{0,\nu} }{2} }
        \int_{ \tau_{0,\nu}/2 }^{\infty} u^{ \dfrac{q}{2} } e^{-u} du \\
        &= \bigg(\dfrac{\tau_{0,\nu}}{2} \bigg)^{- \dfrac{q}{2}} e^{ \dfrac{ \tau_{0,\nu} }{2} } \Gamma \bigg( \dfrac{q}{2}+1, \dfrac{ \tau_{0,\nu} }{2} \bigg)
\end{align}
which is what is presented in the Section \ref{sec:semianalytical}. 

There are two properties of the upper incomplete gamma function $\Gamma(a,x)$. It has an asymptotic behavior for $x \rightarrow \infty$ and fixed $a$: 
\begin{equation}
    \Gamma(a, x) \sim x^{a-1} e^{-x} \bigg[ 1 + \dfrac{a-1}{x} + \dfrac{(a-1)(a-2)}{x^{2}} + ... \bigg]
\end{equation}
and the derivative with respect to $x$ is expressed by: 
\begin{equation}
    \dfrac{ \partial \Gamma (a,x) }{\partial x} = - x^{a-1} e^{-x}
\end{equation}
which we will use to obtain $q_{\text{obs}}$ and $\alpha$. % as presented by Eq. (\ref{eq:qobsp3}) and $\alpha$ in Eq. (\ref{eq:alphap3}). 
% the upper incomplete gamma function approximate the gamma function when x << 1? 

Since $\hat{R_{1}}$ and $\hat{T}_\nu$ can be determined at $\hat{b}=0$ for different wavelengths, we can relate an empirical temperature gradient by Eq. (\ref{eq:qobsdef}) and yield 
\begin{align} \label{eq:qobsp3}
    q_{\text{obs}} = q + \tau_{0,\nu} \Bigg[
            \dfrac{q}{2} + (1 + \tau_{0,\nu}) \bigg[ 
                \dfrac{ \bigg( \dfrac{\tau_{0,\nu}}{2} \bigg)^{q/2} e^{- \tau_{0,\nu} /2 }}{ \Gamma \bigg( \dfrac{q}{2}+1 , \dfrac{\tau_{0,\nu}}{2} \bigg)}
                    - 1 \bigg] 
        \Bigg] \text{ .}
\end{align}
When $\tau_{0,\nu} \ll 1$, the $\Gamma(a,x)$ function in the denominator is a positive finite value of $\sim 1$ and thus one can see that $q_{\text{obs}} \rightarrow q$. Additionally, when $\tau_{0,\nu} \gg 1$, $q_{\text{obs}} \rightarrow q$ also. Numerical calculations show that $q_{\text{obs}}$ slightly deviates from $q$ by $\sim 0.1$ when $\tau_{0,\nu} \sim 1$.

Using the same peak brightness temperature, we can quantify how the spectral index $\alpha$ is related to $q$ in the Rayleigh-Jeans limit. With the spectral index $\alpha$ and the opacity index $\beta$ defined in Section \ref{ssec:spectral_index} and using Eq. (\ref{eq:tbsemi}), we obtain
\begin{align}
    \alpha &= 2 + \beta \dfrac{ \partial \ln \hat{T}_{\nu} |_{\hat{b}=0} }{ \partial \ln \tau_{0,\nu} } \\
        &= 2 + \beta \Bigg[ - \dfrac{q}{2} + \dfrac{\tau_{0,\nu} }{2} - 
            \dfrac{ 
                \bigg(\dfrac{\tau_{0,\nu}}{2} \bigg)^{1+q/2} e^{-\tau_{0,\nu}/2 }
                    }{ \Gamma \bigg( \dfrac{q}{2}+1, \dfrac{\tau_{0,\nu}}{2} \bigg)
                }
            \Bigg] \text{ .} \label{eq:alphap3}
\end{align}
When $\tau_{0,\nu} \gg 1$, we have $\alpha \rightarrow 2 + \beta q (q/2-1) / \tau_{0,\nu}$ which suggests that $\alpha$ is approximately 2 even when there is a temperature gradient. This is mainly because $\hat{R}_{1}$ remains roughly constant at 1 in this limit and one is comparing layers of very similar temperature at different frequencies. Thus, $\alpha$ only retrieves the frequency dependence of the Rayleigh-Jeans tail. When $\tau_{0,\nu} \ll 1$, $\alpha \rightarrow 2 + \beta (-q/2 + \tau_{0,\nu} / 2)$, i.e., $\alpha$ is always less than 2 by $\sim \beta q/2$ given typical values of $\beta$ and $q$ for disks. This relation also demonstrates that the temperature gradient $q$ and opacity index $\beta$ are degenerate in determining the amount of $\alpha$ less than 2. 

\section{Non-Constant Q Disk} \label{sec:nonconstantQ}

In Section \ref{sec:semianalytical}, we have provided a semi-analytical model using a density power index $p=3$. Here we explore other cases where $p=$ 0, 1, and 2 to see the effects. In other words, midplane density profile is not as steep as the constant $Q$ scenario. $Q$ decreases as $R^{p-3}$ and reaches 1 at the edge of the disk. The optical depth along a line of sight presented in Eq. (\ref{eq:optdepth}) is the following for each of the cases:
\begin{align}
    \tau'(\hat{z}) %&= \tau_{0,\nu} \bigg[ \sqrt{1 - \hat{b}^{2}} - \hat{z} \bigg] &\text{for } p = 0 \\
        &= \tau_{0,\nu} \bigg[ \text{arcsinh} \bigg( \dfrac{\sqrt{1 - \hat{b}^{2}}}{|\hat{b}|} \bigg) - \text{arcsinh} \bigg( \dfrac{\hat{z}}{|\hat{b}|} \bigg)  \bigg] &\text{for } p = 1 \\
        &= \dfrac{\tau_{0,\nu}}{\hat{b}} \bigg[ \arctan \bigg( \dfrac{\sqrt{1 - \hat{b}^{2}}}{\hat{b}} \bigg) - \arctan \bigg( \dfrac{\hat{z}}{\hat{b}} \bigg) \bigg] &\text{for } p = 2 
        \text{ .}
\end{align}
Once $\tau'(\hat{z})$ is known, the emergent intensity using the Planck function is 
\begin{equation}
    I_{\nu} (\hat{b}) = \dfrac{2h \nu^{3}}{c^{2}}  
            \int_{-\sqrt{1 - \hat{b}^{2}}}^{\sqrt{1 - \hat{b}^{2}}}
            \dfrac{\tau_{0,\nu}}{(\hat{b}^{2} + \hat{z}^{2})^{p/2}}
            \dfrac{ e^{-\tau'(\hat{z})} }{ \exp{\bigg[ x_{0} (\hat{b}^{2} + \hat{z}^{2})^{q/2} \bigg]} - 1 }
             d\hat{z}
\end{equation}
which corresponds to Eq.~(\ref{eq:I_semi}). Likewise, the brightness temperature in the Rayleigh-Jeans limit and normalized by $T_{0}$ is 
\begin{equation}
    \hat{T} = \int_{-\sqrt{1 - \hat{b}^{2}}}^{\sqrt{1 - \hat{b}^{2}}} \dfrac{\tau_{0,\nu}}{ (\hat{b}^{2} + \hat{z}^{2} )^{(q + p)/2}}
        e^{
            - \tau'(\hat{z})
            }
                d \hat{z}
\end{equation}
which corresponds to Eq.~(\ref{eq:tbsemi}). 

\begin{figure*}
    \centering
    \includegraphics[width=\textwidth]{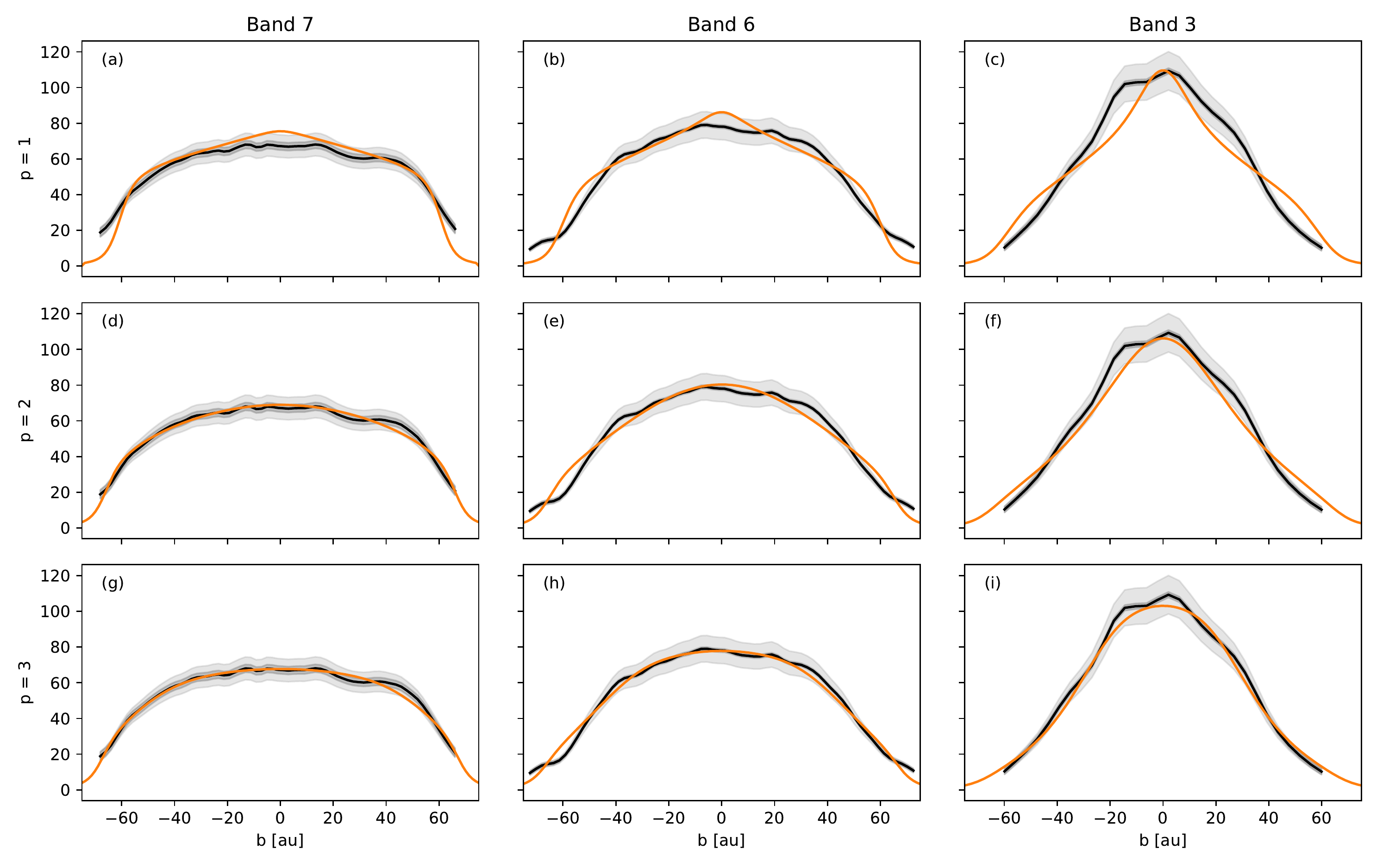}
    \caption{
        The brightness temperature along the major axis at Bands 7, 6, and 3 (from left to right) for different cases of $p$ (from top to down). The black solid lines are the observations with the noise uncertainty plotted as the dark shaded region and the uncertainty including $10\%$ flux calibration uncertainty as the light shaded region. The orange solid line is the best fit model result. 
            }
    \label{fig:pcases}
\end{figure*}

As a demonstration, we fit the Band 7, 6, and 3 profiles using the same procedure in Section \ref{sec:semianalytical} for each of the cases. The fitted parameters are shown in Table \ref{tab:pcases} and the parameters from Table \ref{tab:curvefit} for the $p=3$ case are repeated here for ease of comparison. The resulting profiles are compared to observations in Fig. \ref{fig:pcases}. For the $p=1$ case at Band 6 and 3, there is an evident sharper peak in the center. Since the density is less steep, the $\tau'=1$ surface reaches a smaller radius where the temperature is higher. Towards the side of the profile, the drop is sharper for $p=1$ and more gradual for $p=2$. This is because the contrast of the optical depth near $\hat{b}=1$ and that near $\hat{b}=0$ is smaller for small $p$. For the center to obtain a plateau, the optical depth near $\hat{b}=0$ has to be high and thus the optical depth is also high near $\hat{b}=1$ for small $p$ until it reaches the edge dropping off sharply.

The edge of the disk $R_{0}$ for all three cases lie within 60 to 70 au, since the value is dictated by direct imaging. $T_{0}$ for all cases lie above $45$ K and increases partly because $R_{0}$ decreases for smaller $p$. The temperature power-law index varies from 0.5 to 0.7 for $p$ ranging from 1 to 3. The largest change resides in $\tau_{0,\nu}$, since $\tau_{0,\nu}$ is determined by the shape of the profile and different cases of $p$ produces only certain shapes. Based on the shape of the fits, it is unlikely that $p$ would be as shallow as 1 and fits with $p=2$ and $3$ yield results that differ by no more than a factor of 2.
%it is likely that $p$ is close to 3, if not, in the range of 2 to 3 for HH 212 mms.

\begin{table}
    \centering
    \begin{tabular}{c c c c c}
        $p$     & 1     & 2             & 3 \\
        $R_{0}$ & 61.9 $\pm$ 0.2  & 67.3 $\pm$ 0.3  & 68.7 $\pm$ 0.6\\
        $T_{0}$ & 53.8 $\pm$ 0.6 & 47.3 $\pm$ 0.6  & 45.2 $\pm$ 0.7\\
        $q$     & 0.53 $\pm$ 0.02 & 0.67 $\pm$ 0.02 & 0.77 $\pm$ 0.02 \\
        $\tau_{0,\text{B7}}$ & 1.80 $\pm$ 0.09 & 1.24 $\pm$ 0.06 & 0.90 $\pm$ 0.06 \\
        $\tau_{0,\text{B6}}$ & 1.30 $\pm$ 0.04 & 0.76 $\pm$ 0.02 & 0.52 $\pm$ 0.02 \\
        $\tau_{0,\text{B3}}$ & 0.70 $\pm$ 0.03 & 0.36 $\pm$ 0.02 & 0.20 $\pm$ 0.01 \\

    \end{tabular}
    \caption{Best fit parameters and uncertainties for various cases of $p$}
    \label{tab:pcases}
\end{table}

\section{\bf Effects of Dust Scattering} \label{sec:scattering}

%The intensity calculations across the major and minor axes thus far have not included scattering. In recent years, however, there is increasing evidence for dust scattering through the detection and modeling of polarization at (sub)millimeter wavelengths \citep[e.g][]{Stephens2014, Stephens2017, Bacciotti2018, Kataoka2016}, the measured low spectral index \citep{Carrasco2019, Liu2019, Ueda2020}, or both \citep{Lin2020_specpol}. Therefore, it is beneficial to understand how scattering will affect the interpretations built thus far.

To illustrate the effects of scattering on the distribution of brightness temperature of dust emission, we make use of the 2D axisymmetric disk model described in Section \ref{sec:axisymmetric}, but set the inclination to $90^{\circ}$ (i.e., edge-on) for simplicity. We consider different levels of albedo for a given value of the characteristic extinction (including both absorption and scattering) optical depth $\tau_{0, \nu}$. Albedo at a given wavelength is defined by $\omega \equiv \kappa_{\text{sca}} /  (\kappa_{\text{abs}} + \kappa_{\text{sca}})$ where $\kappa_{\text{sca}}$ is the scattering opacity. Since the extinction optical depth is held fixed, the absorption optical depth decreases while the scattering optical depth increases as the albedo increases. 

\begin{figure*}
    \centering
    \includegraphics[width=\textwidth]{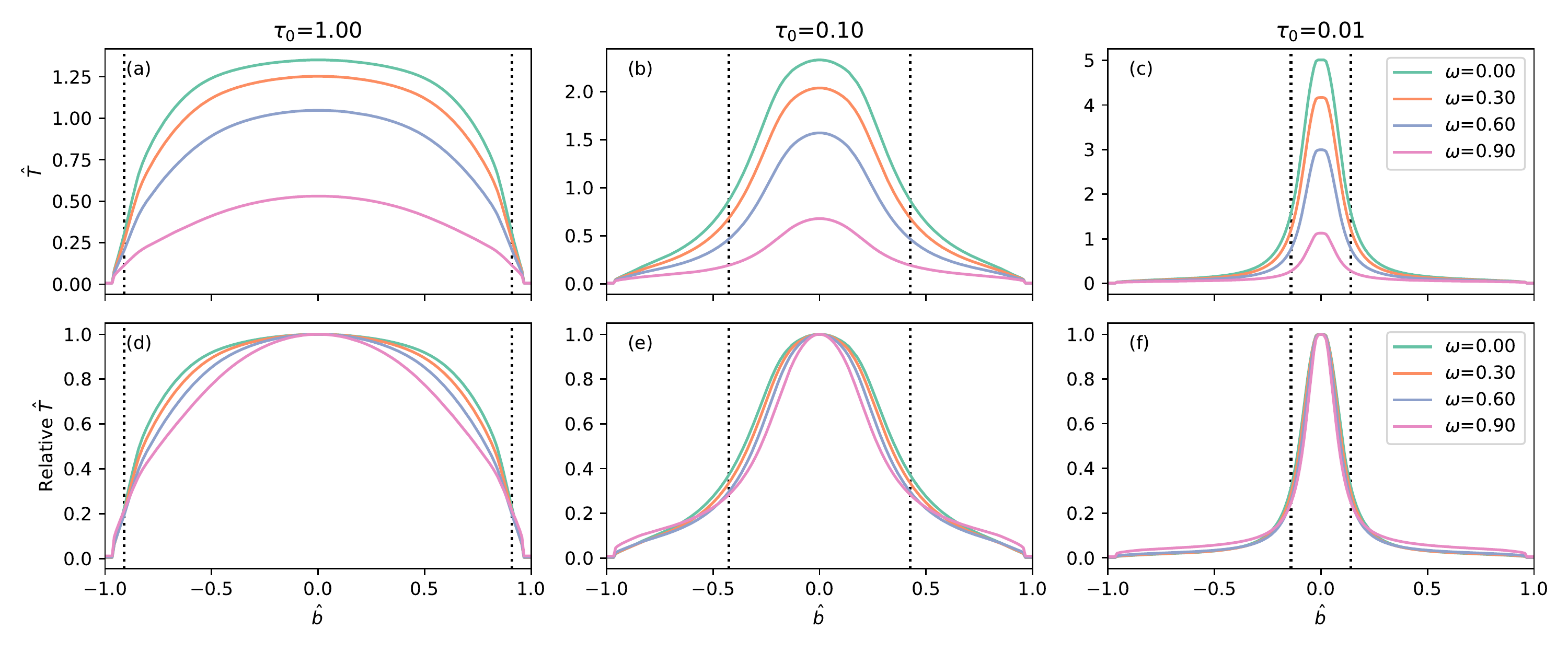}
    \caption{
        Panels (a)-(c): Brightness temperature normalized to $T_{0}$ with different levels of albedo $\omega$ along the normalized impact parameter $\hat{b}$ along the disk major axis. Panels (d)-(f): Each $\hat{T}$ profile is normalized to its own peak to see the relative shape. Each column is varied by the characteristic optical depth $\tau_{0,\nu}$. The black dotted vertical lines mark where the total extinction optical depth becomes 1.
            }
    \label{fig:scat_maj}
\end{figure*}

Fig. \ref{fig:scat_maj}a-c shows the major axis cuts for the model with scattering. For a given $\tau_{0, \nu}$, the brightness temperature decreases with increasing albedo, i.e., with more efficient scattering, the object appears dimmer, %than its actual black body radiation, 
which is in agreement with the plane-parallel calculations \citep{Miyake1993, Birnstiel2018, Zhu2019}. An implication is that, to match the peak brightness temperature observed at a given wavelength, a higher characteristic temperature $T_0$ is needed as the scattering becomes more important. 

% is this part needed???
To compare the shape of the profiles more directly, we normalize each of the cuts by its own peak value and show the results in Fig. \ref{fig:scat_maj}d-f. It is clear that, for a given $\tau_{0,\nu}$, the normalized brightness temperature profile has only a moderate dependence on the albedo: the profile becomes somewhat peakier as the albedo increases. An implication is that, to fit the shape of the brightness temperature profile observed at a given wavelength, a somewhat larger extinction optical depth $\tau_{0,\nu}$ (and thus extinction opacity) is needed as the scattering becomes more important.

\bsp	% typesetting comment
\label{lastpage}
\end{document}